\documentclass[<options>]{elsarticle}
\biboptions{sort&compress}
\usepackage{graphicx}
\usepackage{caption}
\usepackage{threeparttable}
\usepackage{subfigure}
\usepackage{tabularx}
\usepackage{booktabs}
\usepackage{lipsum}
\usepackage{multirow}
\usepackage[usenames,dvipsnames]{color}
\usepackage{colortbl}
\definecolor{mygray}{gray}{.9}

\usepackage{amsmath,bm}
\usepackage[figuresright]{rotating}

\usepackage{algorithm,algpseudocode}
\usepackage{amsmath}
\usepackage{amsfonts}
\usepackage{amssymb}

\usepackage[normalem]{ulem}

\journal{Applied Energy}
\begin{document}
%fmtutil-sys --all
\begin{frontmatter}




\title{On \color{black}Data-Driven Modeling \color{black} and Control in Modern Power Grids Stability: Survey and Perspective\color{black} } 


\author[1,2]{Xun Gong}
\author[1]{Xiaozhe Wang\corref{cor1}}
\author[2]{Bo Cao}
\cortext[cor1]{Corresponding author}
\fntext[cor1]{This work was supported by the Fonds de Recherche du Quebec-Nature et technologies under Grant FRQ-NT PR-298827 and by the Natural Sciences and Engineering Research Council of Canada (NSERC) under Alliance Grants ALLRP ALLRP 571554-21. } %This work was supported by the Fonds de Recherche du Quebec-Nature et technologies under Grant FRQ-NT PR-298827 and FRQ-NT 2023-NOVA-314338.}
%\affiliation[1]{organization={Department of Electrical and Computer Engineering, McGill University},%Department and Organization
 %           addressline={3480 Rue University}, 
 %           city={Montreal},
 %           postcode={H3A 0E9}, 
  %          state={Quebec},
 %           country={Canada}}

\address[1]{Department of Electrical and Computer Engineering, McGill University, 3480 Rue University, Montreal, H3A 0E9, Quebec, Canada}
\address[2]{Huawei Montreal Research Centre, 7101 Park Ave, Montreal, H3N 1X9, Quebec, Canada}         
\begin{abstract}
Modern power grids are fast evolving with the increasing volatile renewable generation, distributed energy resources (DERs) and time-varying operating conditions. The DERs include rooftop photovoltaic (PV), small wind turbines, energy storages, flexible loads, electric vehicles (EVs), etc. The grid control is confronted with low inertia, uncertainty and nonlinearity that challenge the operation security, efficacy and efficiency. The ongoing digitization of power grids provides opportunities to address the challenges with \color{black} data-driven \color{black} and control. This paper provides a comprehensive review of emerging data-driven dynamical \color{black} modeling \color{black} and control methods and their various applications in power grid. Future trends are also discussed based on advances in data-driven control.
\end{abstract}



\begin{keyword}
power grid dynamics and control, data-driven modeling, Koopman operator, data-driven control, physics-informed machine learning, system identification and control


\end{keyword}

%\begin{itemize}
%\item A Koopman-inspired enhanced identification and control is proposed for microgrid secondary voltage and frequency control.

%\item The proposed Koopman-inspired identification requires no knowledge of network information and primary controllers, without warm-up training.

%\item The proposed control is BIBO stable and asymptotically converges under mild conditions.
%\end{itemize}

\end{frontmatter}

\section{Introduction}
Renewable energy is replacing fuel-type generation for sustainable power grids, aiming to reduce greenhouse gas emissions \cite{lund_advances_2019}.
%Microgrids, as an important ingredient of smart grid to aggregate distributed energy resources (DERs), are pursuing zero net energy with 100\% renewable generation.
%Microgrids (MGs) and their evolved forms such as virtual power plant (VPP) and active distribution system (ADS), can be considered as fundamental building blocks where DERs can be aggregated and controlled for secured, effective and efficient grid operation.
%Examples of grid services are grid frequency and voltage control, peak shaving, as well as ancillary services in deregulated electricity markets nowadays. 
The modern power grids are composed of diverse energy resources interconnected through power networks, %Examples are 
which include centralized energy resources %CERs) 
(e.g., synchronous generators, solar and wind farms), and distributed energy resources (DERs) (e.g., distributed renewable generation, energy storage systems, electric vehicles, and thermostatically controlled loads). %The solar and wind energy and 
Particularly, many modernized energy resources are power converter-interfaced and ``low-inertia" in nature. The uncertainty and nonlinearity of them %as well as the interconnection of grid networks 
endanger grid operation security, efficacy and efficiency \cite{lund_advances_2019, Chen2022_RLreviewPS}. 

%\color{black}\emph{uncertainty and nonlinearity of DERs.}\color{black}
The uncertainty and nonlinearity of modern power grids originate from both the energy resources and their systemwide interactions. First, the rapid growth of renewable energy and flexible loads introduces uncertainty and nonlinearity due to the nonlinear stochastic nature of the renewable sources (e.g., wind and solar) and the human behaviors\cite{Chen2022_RLreviewPS}. Second, the interaction dynamics of energy resources through the power networks could suffer from %emerging 
nonlinearity when encountering large disturbances. Besides, the increasing number of diverse energy resources and their hierarchical multi-timescale operation increases the system complexity. All these factors pose challenges to  acquiring accurate system dynamic models and maintaining stable and secure operation of power systems.  %resulting in difficulty in power system operation  difficult and create operational challenges.
%\color{black}\emph{uncertainty and nonlinearity of MG power networks.}\color{black}

Fortunately, the ongoing digitization (e.g., the fast-deploying information, communication and computing techniques) throughout power grids provides opportunities to address the operational challenges %\color{red} \color{black} 
by data-driven control. In this paper, we consider the hierarchical control framework as it is the mature and scalable means to aggregate and manage massive energy resources in modernized power grids %\color{black}
\cite{Olivares2014, MGhierarchy2012, 9281456}. From the bottom to top of the hierarchical structure are distributed energy resources, microgrids/virtual power plants/centralized energy resource, and system operators. The hierarchical control includes primary control at the individual DER level, and secondary and tertiary control at the systemwide level. Examples of the control functions for different control types are summarized in Table \ref{tab:controlfunctions}.

\begin{table} [!b]
\captionsetup{justification=centering, labelsep=newline}
\centering
 \caption{Examples of control functions for different control types }%\cite{MGhierarchy2012}
\label{tab:controlfunctions}
\begin{tabular}{c c}
\hline
Type of Control  & Functions\\
\hline
 & Voltage/frequency stability perseverance,\\
Primary (Component)   & plug-and-play of DERs,\\
 & inertia and local damping control.\\
 \hline
 & Voltage restoration/regulation,\\
Secondary (Area) & wide-area damping control,\\
& frequency regulation.\\
\hline
 & Energy management,\\
Tertiary (Grid)  & optimal operation,\\
  & tie-line power flow control.\\
\hline
\end{tabular}
%\begin{tablenotes}
%\small \item * %$\mathcal{N}(a,b)$ is the normal distribution with mean of $a$ and variance of $b$. 
%Other control parameters are the same to the values in Table I.
%\end{tablenotes}
\end{table}

With the data from the advanced sensing infrastructure (e.g., sensors or transducers, phasor measurement units, smart meters), the energy resources can be coordinately controlled to realize different operation objectives at different levels and time scales in a model-free data-driven fashion.
%\color{black}[data-driven control methods used in modern power grids]\color{black}
To lift the operational challenges, effort has been made to apply data-driven control methods for different use cases in modern power grids and microgrids, such as voltage control \cite{Georgia2021,Ma2021,Ryan2021,Madi2021,Zheng2022}, frequency control \cite{KORDA2018297, Asadi2022,Madani2021,Ryan2021,Madi2021,Zheng2022,Chen2022}, wide-area damping control \cite{Ilias2020,Ilias2018,Guo2021}, cyber-resilient control \cite{Asadi2022,Zhou2021,Liu2021,Chen2022}, as well as demand response \cite{VAZQUEZCANTELI20191072,Liu2019,Mathieu2012,Gong2021}. From the high-level perspective, the data-driven control methods in these studies are based on surrogate models which can be classified into: %\color{red} the categories are not aligned with subsequent sections (done) \color{black} 
(1) linear system data-driven control methods; (2) nonlinear system data-driven control methods. 
%direct data-driven control methods; (2) linear indirect data-driven control methods; (3) pure machine learning-based methods; (4) physics-informed machine learning methods; (5) Koopman-based methods. \color{black}
%(1) universal machine learning methods, and (2) physical-informed data-driven methods. 
This paper provides a comprehensive review on the recent advances of both categories of the methods, with the concentration on the second category that can be further classified into: (2.A) pure machine learning-based methods; (2.B) physics-informed machine learning methods; (2.C) Koopman-based methods. Besides, the paper will discuss the existing applications and trends of these methods in modern power grids.

Note that emerging data-driven methods such as iterative feedback tuning, model-free adaptive control, and learning-based control %(such as iterative feedback tuning, virtual reference feedback tuning, unfalsified control,  model-free adaptive control, learning-based control, safe reinforcement learning, etc) 
are hot topics and reviewed by the researchers in control and robotics communities \cite{hou2016overview,hou2013model, hewing2020learning,brunke2022safe}. For example, the learning-based model-predictive control \cite{hewing2020learning} was investigated for best closed-loop performance through either improved prediction model or proper parameterization of controller (costs and constraints). The mathematical formulation of learning-based control and reinforcement learning (RL) were also discussed based on different safety levels \cite{brunke2022safe}. However, these review papers did not particularly focus on power system applications. %\color{red}Some review papers are from control and other society, not necessarily power systems. But we may also discuss them here and mention they don't particularly focus on power system applications. Check ``An Overview of Dynamic-Linearization-Based Data-Driven Control and Applications'', ``From model-based control to data-driven control: Survey, classification and perspective'', ``Learning-Based Model
%Predictive Control: Toward Safe Learning in Control'', ``Safe Learning in Robotics: From Learning-Based Control to Safe Reinforcement Learning'' (done)
%\color{black} 
The papers \cite{Chen2022_RLreviewPS,Zhang2020_RW} summarized data-driven control in modern power grids, whereas they mainly focused only on the category (2.A) mentioned above,  %\color{red}``first'' may not be clear after my modification. please check (done) \color{black}
i.e., universal machine learning methods such as RL. 
\color{black}
The review paper \cite{She2022} was the first comprehensive summary of current microgrid control framework with applications of RL to address emerging microgrid challenges (i.e., uncertainty and extreme weather).
%\color{red} which still falls in the category of 2.A?  (done)
\color{black} It illustrated the fusion of RL in three ways: (i) model identification and parameter tuning; (ii) supplementary signal generation; (iii) controller substitution. However, the focus of  \cite{She2022} was still pure RL that belongs to the category (2.A); 
%on top of existing microgrid operation framework 
%\color{red} what does "on top of " mean here? (deleted. It was just repeating of the previsous sentense to say that the fusion of RL is discussed under microgrid operation framework) 
the data-driven yet interpretable modeling (with control inputs incorporated) was not illustrated. Besides, emerging data-driven control frameworks other than RL were not investigated.
%whereas details of modeling and identification (with control inputs incorporated), and other emerging data-driven control methods are not investigated.
The review paper \cite{9282004} presented a comprehensive investigation of physics-informed neural network (NN) for power system applications, which fell in the category (2.B) whereas the review concerned only a specific learning machine (i.e., NN) and the incorporation of control was not discussed in detail. 

Motivated by the limitations of previous review papers, the goal of this paper is to provide a comprehensive method review of data-driven approaches with a broader scope, aiming to %inspire reliable data-driven solutions to 
address increasing time-varying uncertainty and nonlinearity in modern power grids.  It is important to note that data-driven control methods are not intended to replace existing model-based control frameworks \cite{She2022}, but rather to complement them as supplementary or enhancement solutions driven by data. %These methods should seamlessly integrate with existing model-based controllers in a manner that is explainable and scalable. 
By offering a broader technology map and comparative vision of different data-driven control approaches, this paper aims to inspire new ideas and feasible solutions in power systems. %Considering practicality, we consider that data-driven control methods are not designed to fully replace existing control frameworks and model-based control \cite{She2022}. Rather, they are supplementary or enhancement solutions driven by data, and ultimately capable of seamlessly integrating with existing model-based controllers in an explainable and scalable fashion. 
%The paper seeks to provide a more encompassing technology map and a comparative vision of various data-driven control approaches, with the aim of inspiring novel ideas and feasible solutions within the power sector. In comparison to previous review works \cite{Zhang2020_RW,Chen2022_RLreviewPS,9282004}, this paper introduces the following key contributions:
%We hope to provide a more general technology map and comparative vision of different data-driven control approaches, aiming to inspire new ideas and feasible data-driven solutions in power sectors. 
Compared to previous review works \cite{Zhang2020_RW,Chen2022_RLreviewPS,9282004}, this paper makes the following key contributions: %the main contributions of this paper are:

(1) It includes a broader range of methods, encompassing linear identification and control techniques based on input-output models, state space representations, transfer function identification, as well as nonlinear methods such as reinforcement learning, supervised learning-based and Koopman-based approaches. 
%Some other methods (including a bunch of linear identification and control based on input-output models, state space representations and transfer function identification, as well as nonlinear methods such as supervised learning-based methods and Koopman-based methods, etc.) are also included.

(2) It provides a comprehensive comparison of data-driven approaches from various perspectives, including modeling structure, identification and control methods, data requirements, adaptiveness, interpretability, scalability, and training efficiency. 
%The data-driven approaches are comprehensively compared with each other from different perspectives such as modeling structure, identification and control methods, data requirements, adaptiveness, interpretability, scalability, training efficiency, etc. 
Particularly, this paper introduces a notable advancement by including and reviewing Koopman-based methods for the first time. These methods are categorized as (2.C) and are particularly promising in addressing challenges such as nonlinearity and uncertainty while leveraging established linear identification and control techniques in conjunction with emerging machine learning methodologies. %Among all the reviewed approaches, Koopman-based methods (category 2.C) are firstly reviewed and promising in terms of addressing issues in new applications on top of old methods, i.e, using mature linear identification and control methods to solve nonlinearity and uncertainty with the help of emerging machine learning techniques. 
% 
%\color{red} a little bit hesitant to include the next sentence or not (I commented out and it looks fine)\color{black}
%This bridges the gap between linear and nonlinear methods, and thus rich mature theories and techniques can be adapted to help address unsolved issues including method interpretability, safety and scalability of complex modern power grids, training efficiency, etc.

%(3) The review paper focuses more on data-driven methodology on top of which the practicality of power grid applications are considered. The paper is well suited for inspiring reliable solutions that are better suited to incorporate control input channels in power system modeling and control stages.

(3) The paper organizes different categories of data-driven methods systematically to generalize the data-driven control frameworks, enabling researchers and practitioners to navigate through the diverse landscape of data-driven approaches in a more systematic manner, fostering novel ideas and feasible data-driven solutions in power systems. 

%\color{red} Do you feel that you provide a unified framework to put different categories of methods together? Or do you feel that the paper specifically tailors power system applications? (done) \color{black}

%Thus, we hope to provide a broad vision to fill the gap, aiming to provide researchers and engineers in power sectors a more general technology map of data-driven control and the potential for grid applications.
%\color{red} How this review is different from existing review papers; why we need another review (done)
\color{black}
 
%\color{black} 
%The rest of this paper is organized as follows. Section 2 presents the preliminaries and classification of state-of-the-art data-driven control methods. Sections 3-4 present critical reviews for both categories of data-driven methods. Specifically, Section 3 elaborates on the linear system data-driven control methods. Section 4 details nonlinear system data-driven control methods. Section 5 provides an overview and discusses the key safety issues of existing methods in power grid applications. Section 6 concludes the paper. %\color{black} 

%\color{black} 
The rest of this paper is organized as follows. Section 2 presents the preliminaries and a brief history of state-of-the-art data-driven control methods in power grid applications. Sections 3-4 present critical reviews for both categories of data-driven methods. Specifically, Section 3 elaborates on the linear system data-driven control methods. Section 4 details nonlinear system data-driven control methods. Section 5 provides a grid application overview and discusses the future trend. Section 6 concludes the paper. %\color{black} 

%Section II presents the preliminaries and review of state-of-the-art data-driven control methods in general from the theoretical perspective. Sections III-IV present the state-of-the-art applications of linear data-driven control methods in power grids. \color{red} check out how we wrote the abstract and outlines and modify this part. \color{black} Section V discussed the state-of-the-art applications of nonlinear data-driven control methods in power grids. Section VI concludes the paper.

%The rest of this paper is organized as follows. Section II presents the preliminaries and review of state-of-the-art data-driven control methods in general from the theoretical perspective. Section III presents the state-of-the-art applications of linear data-driven control methods in power grids. Section I.V discussed the state-of-the-art applications of nonlinear data-driven control methods in power grids. Section V concludes the paper.
%\emph{Voltage Control.}
%\emph{Frequency Control.}
%\emph{Wide-Area Damping Control.}
%%\emph{Frequency Regulation.}
%\emph{Cyber-Resilient Control}
%\emph{*Demand Response.}

%[What has been solved and what remains]

%to provide the services to the bulk grids connected through MG energy routers.
%Flexible loads can adjust the active power for some ancillary services at relatively slow time scales (e.g., regulation, load following, spinning reserves, etc.).

\section{\color{black} Preliminaries and an Overview on Data-Driven Control for Power System Applications}
%\color{black}[Advances and trends of general data-driven control methodology]\color{black}
\color{black} This section presents an overview of data-driven identification and control methods and their applications in power systems. First of all, the following terminologies adopted in the manuscript are synonyms of or highly related to data-driven control, thus they need to be clarified to avoid conceptual ambiguity. 

\textbf{1) Data-driven control}: the term ``data-driven" means that the identification and/or the design of the controller are based entirely on experimental data collected from the plant or simulators but not any explicit information from first principle mathematical models of the controlled process \cite{hou2013model}. Data-driven control often refers to the closed-loop control starting and ending up with data \cite{hou2013model}. In this paper, we focus on the \textbf{\emph{data-driven control that is based on data-driven modeling}}, which is defined below.

\textbf{2) Data-driven modeling}: the modeling method based on identification or learning from data while disregarding explicit knowledge of the system’s physical behavior \cite{zarkogianni2019personal}. The modeling and identification/learning can be done with or without control inputs incorporated.

\textbf{3) Model-free control}: the \emph{``model"} in \emph{model-free control} and \emph{model-based control} refers to the first principle physical model of the system of interest. The term ``model-free" refers to an alternative technique to control systems without traditional first principle physical models. This can be done by using a simplified representation of the system in a data-driven fashion. In other words, it is a synonym for data-driven control \cite{hou2013model}.

%\textbf{4) Adaptive control}: \emph{adaptive control is the control method used by a controller which must adapt to a controlled system with parameters which vary, or are initially uncertain} \cite{egardt1979stability}. The adaptive control itself can be conducted based on either physical parametric models or data-driven models or model-free representations). For data-driven models, the foundation of adaptive control is online real-time parameter estimation for adaptive identification \cite{shakeel2020line, wu2006multivariable}. Another way to realize adaptive data-driven control (without an explicit data-driven modeling stage) is to tune the controller parameters according to past or simulation data (e.g., tuning the fuzzy PI controller with heuristic optimization \cite{keshta2021fuzzy}). Generally speaking, data-driven control can realize adaptiveness in two respects: (i) offline identification/learning can help deeply learn the inherent problem structure when sufficient data is collected, whereby slow adaptiveness can be achieved. (ii) Online identification/learning can quickly adjust the model parameters to achieve fast adaption to time-varying uncertainty. \color{red} TBD\color{black}

\textbf{4) Adaptive control}: {adaptive control is the control method used by a controller which must adapt to a controlled system with parameters which vary, or are initially uncertain} \cite{egardt1979stability}. The adaptive identification and control itself can be conducted based on either physical parametric models or data-driven models (model-free representations). Data-driven control may realize adaptiveness through online identification/learning that quickly adjusts the model parameters to achieve fast adaption to time-varying uncertainty. %\color{red} how about keeping this sentence? (sure) \color{black} 
%Generally speaking, data-driven control can realize adaptiveness in two respects: (i) offline identification/learning can help deeply learn the inherent problem structure when sufficient data is collected, whereby slow adaptiveness can be achieved. (ii) online identification/learning can quickly adjust the model parameters to achieve fast adaption to time-varying uncertainty. %For data-driven models, the foundation of adaptive control is online real-time parameter estimation for adaptive identification \cite{shakeel2020line, wu2006multivariable}. They 
%\color{red} TBD (last sentence deleted) \color{black}

%\textbf{5) Offline/Online training}:  The offline training refers to the training of learning machines  (such as neural networks) to obtain optimal parameters (weights and biases) with offline data (i.e., the datasets stored offline). On the contrary, online learning means that the training is conducted based on the data collected online and the learning machines are trained on a rolling basis.

%\textbf{6) Supervised learning}: The word "supervised" means a machine learning paradigm for problems where the available data consists of labeled examples whereby the learning can be conducted under supervision of the label. 

According to the above-mentioned context:
(i) Model-free control and data-driven control are mainly the same concepts that are interpreted from different perspectives of a method. Although certain physics-inspired information (such as structure design or constraints) may be incorporated in data-driven modeling and control, we assume model-free control and data-driven control are interchangeable terms in the manuscript.  
(ii) Data-driven control can be adaptive or not depending on whether online identification is realized. \color{black}
The two assumptions apply to the rest of the paper without further explanation.

Generally, the data-driven control methodologies can be classified by the assumed \color{black} data-driven model\color{black}, i.e.,  linear system control and nonlinear system control. \color{black} In power system applications, the linear transfer function model and identification were first investigated since early-nineties to model the dynamical modes of power grids. An example is the transfer function of impulse response between small-signal inputs and outputs with least-squares identification presented in 1993 \cite{Smith1993}. The identified transfer function can be applied for different purposes such as: 
(i) tuning, design, and testing of power system control systems such as power system stabilizer (PSS), static var compensator (SVC) and many others;
(ii) validation of power system small-signal models for grid planning and operation. %the accurately identified transfer function with small signals, as a good small-signal representation of power grid, are also useful for power system operators to validate model and simulation software used for grid planning and operation. 
Different transfer functions can be identified to deal with different potential operating conditions, whereby robust controllers can be designed accordingly.
%However, the transfer function identification relies on properly selected inputs and outputs (e.g., small signals of interest for ambient conditions) and the associated assumption that the input-output system is an LTI (linear time-invariant). 
%As power grids are nonlinear by nature, the method is inherently not suitable for large-signal  scenarios where the nonlinearity and uncertainty emerge (e.g.,transients/faults of power systems, dynamics of low-inertia power grids subject to large disturbance and increasing volatile renewables). 
%The data also matters. Once the signals used for identification is introducing nonlinearity (e.g., the control signals get saturated). then the identification accuracy is compromised or even leads to ineffective modeling \cite{Smith1993}.
In the meantime, the authors in \cite{kamwa1993minimal} proposed a reduced-order linear state space model based on a minimal realization approach for modal analysis,  
%with transient stability data, 
which shows the effectiveness of linear methods identified by transient (sometimes termed as ring-down \cite{sanchez2012identification}) data  for conventional power systems with large rotating masses. Similar work on multiple-input multiple-output (MIMO) state space identification based on pulses can be found in \cite{kamwa2000state}, which was effectively applied to the modal analysis of bulk power systems. The subspace methods including ORT, N4SID, MOESP, CVA) were used for the linear state space identification with different probing tests \cite{kamwa2000state,Kamwa1995,Zhang2013mimo}. 
Afterwards, many variants of transfer function identification methods \cite{gurrala2015loewner,rergis2018loewner, Liu2017,ramakrishna2010adaptive, Zelaya2020,8882329,8973540,Barkley2009,zhang2015data,YangDW2020,Madani2021,Ryan2021} and state space identification methods \cite{WideAreaControl_N4SID,6082169,Mathieu2012,Nieto2015,Zhang2016,Ilias2020,Georgia2021} were developed for data-driven control in the context of power grid applications such as damping control, voltage control and microgrid control (primary, secondary and tertiary) and aggregated load control.
Generally, 
both transfer function and linear state space methods can represent systems well using a locally linearized model around a fixed operating point under the assumption that the system is an LTI (linear time-invariant). %but
%state space representations are typically less cumbersome than transfer functions (polynomial representations) in multivariate cases \cite{wu2006multivariable}. %Both transfer function and linear state space methods can represent systems well using locally linearized models around fixed operating points. %the state space representations are less cumbersome than transfer functions (i.e., polynomial representations) in multivariate cases \cite{wu2006multivariable}. 
%What is common among the transfer function and linear state space methods is that the systems can be well represented with locally linearized models around the fixed operating points. 
%Also, they rely on the assumption that the system is an LTI (linear time-invariant). However, the assumption of a local linearized model does not hold when the system is subject to large disturbances. %for large-signal disturbance such as power system transients that incur nonlinear and non-stationary dynamics. 
%Besides, 
However, modern grids tend to be low-inertia with the increasing penetration of volatile inverter-interfaced renewables, leading to wide-range dynamics with higher levels of time-varying uncertainty and nonlinearity. This compromises the LTI assumption. Also, the saturation of control signals and nonlinearity of control input channels can also undermine the LTI assumption, leading to a decrease in identification accuracy. %may also compromise the assumption of LTI thus  %Besides, when the control signals used for identification is introducing nonlinearity (e.g., the control signals get saturated), the assumption of LTI is compromised, thus 
%decreasing the identification accuracy.  % or even leading to ineffective modeling 
%\cite{Smith1993}. %\color{red} is the last sentence used to explain data matters? (modified) \color{black} 
The details of different linear identification and control methods will be discussed in Section 3.

Nonlinear identification methods are suited to address the above-mentioned challenges. Traditionally, nonlinear identification methods based on Hibert-Huang Transform (empirical mode decomposition + Hilbert analysis) are used to adaptively model power system nonlinear dynamics \cite{sanchez2012identification}. Supervised learning (i.e., a main branch of machine learning) based nonlinear modeling started emerging in the mid-nineties because of the more powerful nonlinear fitting capability of universal learning machines such as NNs \cite{mori1992artificial,Kamwa1996RNN,ku1994power,bostanci1997identification}. For example, in 1996, Innocent Kamwa et al \cite{Kamwa1996RNN} proposed a MIMO supervised-learning recurrent neural network (RNN) that is equivalent to a general differential equation, which can be trained offline with the output-layer parameters then identified online in a recursive fashion to adaptively describe time-varying nonlinear power system dynamics.
However, supervised learning is more often used for nonlinear dynamic modeling without external control. 
To incorporate nonlinear control input channels in the modeling, adequate training data are necessary, which could be generated by actively probing test signals then collecting the response data for a long period of time. Such requirement is not practical as it is often not allowed to subjectively inject large signals into safety-critical systems like power grids, while probing low-level test signals may not sufficiently excite the system to obtain informative data. %\color{red}\cite{}\color{black}. 
%\color{red} I think we may make the structure clearer. This paragraph talks about supervised-learning (needs to be improved, may make it clear at the beginning of the paragraph), and the next paragraph focuses on reinforcement learning (clear enough) [done. Transfer learning part is moved to Section 4.1 Page 18-19 in red]\color{black}

Reinforcement learning (RL), as another branch of machine learning that inherently bridges control and mainly rely on offline simulators, becomes popular nowadays for the applications of data-driven control of power grids \cite{Chen2022_RLreviewPS,Zhang2020_RW,She2022}. 
%The learning process generally depends on offline simulators rather than probing data to real systems \cite{Chen2022_RLreviewPS}. 
By interacting with the simulation environment to pursue Bellman optimality, the RL can yield optimized control policy that can be deployed. The control optimality is theoretically sound while without performance guarantee due to the existence of simulator uncertainty/modeling error and optimization error \cite{haykin_neural_2010}. Combining RL with deep learning (i.e., deep reinforcement learning (DRL)) becomes popular to enhance mathematical optimality and generalization capacity of learning, and shows the effectiveness in power grid applications such as optimal voltage control \cite{KOU2020114772}, frequency regulation \cite{Khooban2021}, EV charging scheduling \cite{Wan2019}, and battery management \cite{Bui2020}. To improve the efficiency of deep learning , DRL may be developed under parallel computing frameworks which has been shown effective in autonomous voltage control and emergency control \cite{Huang2020, Xu2020, HuangR2022}.
%such as generator dynamic braking, under-voltage load shedding \cite{Huang2020, Xu2020, HuangR2022}.
The details of pure machine learning-based identification and control methods, including supervised and reinforcement learning, will be provided in Section 4.1.

Although pure machine learning methods stated in the last two paragraphs are of full capacity to fit any nonlinearity, their practical applications in power systems face pressing challenges such as %there are a few challenges (such as 
physical consistency, interpretability,  generalization, safety, etc.  %training efficiency, data adequacy and safety) 
Physics-informed rules and laws (physics-informed loss function, constraints, initialization, architecture design, hybrid and ensemble learning, etc.) become a growing consensus to mitigate the challenges \cite{Huang2022RW}. %to %reduce uncertainty in data-driven modeling and 
%enhance the physical consistency, generalization, and interpretability of learning machines. \color{red} please dou
%in terms of reducing data-driven modeling uncertainty and make the adopted learning machines more physically consistent,  generalized and interpretable. 
The fusion of RL frameworks and physics-inspired information is another promising solution as it leverages the intrinsic connection between machine learning and control theories. %that needs attention for data-driven control, when considering the intrinsic connection between machine learning and control theories. 
The fusion can be conducted through catering the physics-informed rules and laws of supervised/unsupervised learning for RL in environment surrogate model, value function or policy design \cite{Huang2022RW, Gros2020,ghavamzadeh2015bayesian,brunke2022safe}. %Nonetheless, the structural interpretability and the way physics are informed can still be enhanced to reduce the possibility to obtain physical-inconsistent solutions that may deteriorate data-driven control performance or even destabilize power grids. Besides, reinforcement learning based methods still suffers from some practical issues for power grid applications, such as the lack of scalability, adaptiveness and training efficiency \cite{She2022}. 
However, further improvements are necessary to enhance structural interpretability and avoid physically inconsistent solutions, which could compromise data-driven control performance or even destabilize power grids. Additionally, machine learning based methods still face practical challenges in power grid applications, including data availability, training efficiency, scalability, and adaptiveness. \cite{She2022}. The discussion about physics-informed machine learning-based methods will be given in Section 4.2. \color{black}

%To improve the efficiency and adaptiveness, the authors in \cite{Huang2020, Xu2020, HuangR2022} proposed DRL with parallel computing frameworks 
%\color{red}in (what applications)\color{black}. \color{red}I would give more general discussion here while moving the detailed discussion to Section 4. \color{black}
%The parallel computing can facilitate the multi-task training of deep reinforcement learning. A wide variety of scenarios in power grids can be created in simulation to train DRL. After trained, DRL possesses certain level of generalization capacity and can be used for some new scenarios. However, such adaptiveness is limited as it highly depends on the simulator data quality, and the system operating conditions and topology may change over time and may be not known by system operators. The old simulation data may not be able to reflect the underlying change of physical systems; thus the fast adaptation to new operating conditions is compromised. The details of nonlinear physics-informed machine learning methods will be provided in Section 4.2.

The unsolved issues further motivate recent research works on adaptive Koopman operator control with online nonlinear identification \cite{Gong2022,gong2023novel}, aiming to adaptively map nonlinear control to linear control that works for both small and large signals. Specifically, the methods adopt small-signal linear space augmented with more physically interpretable nonlinear bases and are adaptively identified to address time-varying uncertainty. Although these methods are online and can be applied without warm-up training, Koopman state space is still determined empirically. Koopman generators that can estimate optimal and physical-consistent Koopman operators (based on physical states and control inputs) could be exploited with physics-informed learning and adequate offline data. The details of Koopman-based methods will be provided in Section 4.3.

%\color{red} Is it possible to relate the history to the contents to be discussed in different sections? (a sentence was added at the end of each paragraph to refer readers to section 3 and 4.)
\color{black}

\color{black}
In the following, we will concentrate on elaborating existing data-driven control methodologies in linear systems (Section 3) and nonlinear systems (Section 4) used in power grids. The overview of grid applications with these methods is provided in Section 5.
\color{black}

\section{ Linear System Identification and  Control}
\color{black}
%\textbf{\emph{Linear data-driven control in power systems}}. For small signals of nonlinear power systems, the nonlinearity can be well described by the linearized small-signal model structure, whereby linear data-driven control can be applied. Examples of the power system applications include but are not limited to wide-area damping control \cite{Guo2021,Ilias2018,Ilias2020}, voltage control \cite{Pierrou2021}, microgrid secondary control \cite{Madani2021}, microgrid tertiary control \cite{Zhang2016}, primary converter control \cite{HUANG2021192,Ryan2021}, decentralized damping control \cite{Huang2022}, aggregated load control \cite{Mathieu2012}, etc.
\color{black}

In existing data-driven control methodologies for linear systems, identification and control are typically regarded as two tasks that are done in a sequential manner. Specifically, when the data of inputs and outputs (i.e., $\bm{u}_d$ and $\bm{y}_d$) are available, the identification of a model of interest $\bm{y}=\hat{\bm{y}}(\bm{x},\bm{u};\bm{\theta})$ for the unknown parameter $\bm{\theta}$ and state $\bm{x}$ can be written in a general form of minimizing a loss function $\mathcal{L}_{id}$ as below:
\begin{small}
\begin{subequations}
\begin{equation}
\bm{\hat{\theta}} = arg \min_{\bm{\theta},\bm{x}}\mathcal{L}_{id}(\bm{y}_d,\hat{\bm{y}}(\bm{x},\bm{u}_d;\bm{\theta}))
%=  arg\min_{\theta} \frac{1}{N}(y - \hat{y}(t|\theta))^2
%\bm{x}_k+\bm{B}\bm{u}_k+\bm{\delta}_k
\label{eq:id_general}\end{equation}
\begin{equation}
\bm{x} \in \mathcal{X}, \quad
\hat{\bm{y}} \in \mathcal{Y}
\label{eq:u_constraint}\end{equation}
\end{subequations}
\end{small}\noindent
%$\hat{y}$ represents the system model which is a function of $u$ with the parameter $\theta$. 
where $\mathcal{X}$ and $\mathcal{Y}$ are the state constraint set and the output constraint set, respectively.  The control thereafter can be applied to the identified model with $\hat{\bm{\theta}}$, which is equivalent to the optimization in a general form of
\begin{small}
\begin{subequations}
\begin{equation}
%\{\bm{\hat{u}}\} = arg
\min_{\bm{u},\bm{x}}\mathcal{L}_{c}(\bm{y}_r,\hat{\bm{y}}(\bm{x},\bm{u};\hat{\bm \theta}),\bm{x})
%=  arg\min_{\theta} \frac{1}{N}(y - \hat{y}(t|\theta))^2
%\bm{x}_k+\bm{B}\bm{u}_k+\bm{\delta}_k
\label{eq:ctrl}\end{equation}
\begin{equation}
\bm{u} \in \mathcal{U}, \quad
\bm{x} \in \mathcal{X}, \quad
\hat{\bm{y}} \in \mathcal{Y}
\label{eq:uy_constraint_c}\end{equation}
\end{subequations}
\end{small}\noindent
where $\bm{y}_r$ is the reference output desired by control;  $\mathcal{U}$ is the input constraint set; $\mathcal{L}_{c}$ is the loss function representing the control objectives in general.

Alternatively, the identification and control can be done simultaneously and directly with respect to the input $\bm{u}$ and output $\bm{y}$ while without the state $\bm{x}$. That is
\begin{small}
\begin{subequations}
\begin{equation}
%\{\bm{\hat{\theta}}, \bm{\hat{u}}\} = arg 
\min_{\bm{\theta},\bm{u},\bm{y}}\{\mathcal{L}_{id}(\bm{y}_d,\bm{u}_d,{\bm{f}}(\bm{y},\bm{u};\bm{\theta}))+\alpha\mathcal{L}_{c}(\bm{y}_r,{\bm{f}}(\bm{y,u;{\theta}})) \}
%=  arg\min_{\theta} \frac{1}{N}(y - \hat{y}(t|\theta))^2
%\bm{x}_k+\bm{B}\bm{u}_k+\bm{\delta}_k
\label{eq:id_ctrl}\end{equation}
\begin{equation}
\bm{u} \in \mathcal{U}, \quad
\hat{\bm{y}} \in \mathcal{Y}
\label{eq:uy_constraint_direct}\end{equation}
\end{subequations}
\end{small}\noindent
where $\alpha$ is the coefficient to weight the identification and control objectives in (\ref{eq:id_ctrl}). In short, the identification-control tasks in sequential and simultaneous fashions  correspond to the indirect and direct data-driven control in the control community. Interested readers can refer to \cite{Dorfler2022} for more theoretical details. In the following, we will discuss the %tailored %identification and control task formulation in what follows 
indirect and direct linear system data-driven control methods in existing and potential power system applications, respectively.
%In the following, we will detail the problem formulations for power system applications with different data-driven control methods.
%Specifically, two main categories of data-driven control and their state of the arts (i.e. linear control and nonlinear control) are illustrated. In each category, different data-driven techniques have been applied to realize effective identification and control. 
%\color{red} may need to give the system model (state-space model) \color{black}
\subsection{{Sequential Linear Identification and Control}}
For sequential identification and control, the identification plays an important role to realize effective modeling and thus control. In what follows, we focus on linear identification techniques that can be categorized as 
\emph{state space model-based} and \emph{input-output model-based}. The state space model is generally more suitable for multi-variable system modeling as it deals with the individual input/output variables in a vector space. Besides,  intermediate states properly selected can help define the inherent input-output relationship when compared to the input-output ``black-box" modeling based on transfer functions. %\color{red} not very clear at this stage without seeing the detailed formulation. TBD\color{black}

\subsubsection{{Linear System Identification Based on State Space Model}}
%Conventionally, 
We consider a special form of the model representation $\hat{\bm{y}}(\bm{x},\bm{u};\bm{\theta})$ in (\ref{eq:id_general})
%When power systems operate under ambient conditions, the nonlinear power system can be well described by the linearized small-signal model, which can be represented in a state space form \cite{lange_physical_2006, Ilias2020}. %\color{black} \color{red} can you relate (4) to (1a)? I think (4) is a special form of function L \color{black}
\begin{small}
\begin{subequations}
\begin{equation}
%(Process \quad  model) 
\bm{x}_{k+1} = \bm{A}\bm{x}_k+\bm{B}\bm{u}_k+\bm{\delta}_k
\label{eq:process}\end{equation}
\begin{equation}
\bm{y}_k = \bm{C}\bm{x}_k + \bm{D}\bm{u}_k+\bm{e}_k%+\bm{D}\bm{u}_k+\bm{e}_k
\label{eq:ssmodel}\end{equation}
\end{subequations}
\end{small}\noindent
which is termed as the state-space model and has been widely used to represent power systems dynamics operating in ambient conditions. %the nonlinear power system can be well described by the linearized small-signal model, which can be represented in a state space form
%The equations (\ref{eq:process})-(\ref{eq:ssmodel}) is a special form of the model representation $\hat{\bm{y}}(\bm{x},\bm{u};\bm{\theta})$ in (\ref{eq:id_general}), \color{black} where 
$\bm{A}$, $\bm{B}$ and $\bm{C}$ are the parameters corresponding to $\bm\theta$. %in (\ref{eq:id_general}) or (\ref{eq:id_ctrl}). 
$\bm{x}_k$ are the states of power systems (e.g., phasor angle, rotor angle, frequency, voltage), $\bm{y}_k$ are the observation outputs of power systems (e.g., frequency, voltage, power), and $\bm{u}_k$ are the control inputs (e.g., the reference power of generators). $\bm{\delta}_k$ and $\bm{e}_k$ are the process noise and the observation noise, respectively. The linear state space form is conducive to estimation, filtering, prediction and control. 
When system parameters are unknown, mature linear identification and control techniques can be used to apply on the system (\ref{eq:process})-(\ref{eq:ssmodel}). 
%The identification and control be implemented either sequentially or simultaneously. 

%\textbf{\emph{Classical linear system identification and control}.}

%\subsubsection{{Classical Linear System Identification based on State Space Model}}%Typically, the identification and control tasks are implemented sequentially. 
%The key is to realize effective identification whereby mature linear controllers can be applied to the identified model. 
Given the system model in (\ref{eq:process})-(\ref{eq:ssmodel}), classical linear identification can be applied to identify the parameters $A$, $B$, and $C$. %the model.  
%\cite{SURANA2016716, Netto2018,BAI20161}
%Least-squares estimation \cite{SURANA2016716,Netto2018} can be applied directly based on (\ref{eq:process})-(\ref{eq:ssmodel}). 
%\color{black}For example, extended Kalman filter can be applied based on (\ref{eq:process})-(\ref{eq:ssmodel}) by treating the parameters as the "states" to estimate \cite{Mathieu2012}.
%\color{red} are Kalman filter methods model-based?  %\color{black} (A: the model parameters are unknown in this case so I think this is data-driven.)
Examples of classical linear system identification methods are subspace methods including CVA \cite{Ghasemi2006}, N4SID\cite{Ghasemi2006,Jamaludin2013,Mansouri2018,WideAreaControl_N4SID,Yohanandhan2016,Nieto2015,Zhu2016,dosiek2012mode}, MOESP \cite{Ye2010,Jamaludin2013}, OKID \cite{alenany_modified_2019}.
Generally, the data matrices with Hankel structure play an important role in these subspace methods because the signals (the input data, output data and the noises) in these algorithms are organized in the form of Hankel matrices \cite{katayama2005subspace, QIN20061502}. Specifically, let $\bm{U}_P$, $\bm{Y}_P$, $\bm{U}_F$ and $\bm{Y}_F$ represent the ``past" input and output data, as well as the ``future" input and output data, respectively. $[\bm{U}_P,\bm{U}_F]^T$ constitutes the Hankel matrix of the input $\bm{u}$, and $[\bm{Y}_P,\bm{Y}_F]^T$ constitutes the Hankel matrix of the output $\bm{y}$. The Hankel matrices with the depth of $L$ and trajectory length $T$ are as follows: 
\begin{small}
\begin{subequations}
%\begin{aligned}
\begin{equation}
\begin{bmatrix}
\bm{U}_P\\
-\\
\bm{U}_F\end{bmatrix}
=
\begin{bmatrix}
\bm{u}_0 & \bm{u}_1  & ...& \bm{u}_{T-L}\\
\bm{u}_1 & \bm{u}_2 & ...&\bm{u}_{T-L+1}\\
\vdots & \vdots & \ddots & \vdots \\
\bm{u}_{L-1} & \bm{u}_{L} & \dots &\bm{u}_{T-1}
\end{bmatrix}\label{eq:HankelU}
\end{equation} 
\begin{equation}
\begin{bmatrix}
\bm{Y}_P\\
-\\
\bm{Y}_F\end{bmatrix}
=
\begin{bmatrix}
\bm{y}_0 & \bm{y}_1  & ...& \bm{y}_{T-L}\\
\bm{y}_1 & \bm{y}_2 & ...&\bm{y}_{T-L+1}\\
\vdots & \vdots & \ddots & \vdots \\
\bm{y}_{L-1} & \bm{y}_{L} & \dots &\bm{y}_{T-1}
\end{bmatrix}\label{eq:HankelY}
\end{equation} 
\end{subequations}
\end{small}\noindent
The data matrices can be arranged according to (\ref{eq:process})-(\ref{eq:ssmodel}) as:
\begin{small}
\begin{subequations}
\begin{equation}
%(Process \quad  model) 
\bm{Y}_{P} = \bm{\mathcal{O}}_{L/2}\bm{X}_P+\bm{\mathcal{H}}_{L/2}\bm{U}_P
\label{eq:datamatrixSubspace1}\end{equation}
\begin{equation}
\bm{Y}_{F} = \bm{\mathcal{O}}_{L/2}\bm{X}_F+\bm{\mathcal{H}}_{L/2}\bm{U}_F
\label{eq:datamatrixSubspace2}\end{equation}
\end{subequations}
\end{small}\noindent
where $\bm{\mathcal{O}}_{L/2}=[\bm{C},\bm{CA},\dots,\bm{CA}^{\frac{L}{2}-1}]^T$ is the extended observability matrix; $\bm{H}_{L/2}$ is the lower Toeplitz matrix defined as:
\begin{small}
%\begin{aligned}
\begin{equation}
\bm{\mathcal{H}}_{L/2} =
\begin{bmatrix}
\bm{D} & 0  & ...& ...& 0\\
\bm{CB} & \bm{D} & ...& ...& 0\\
\bm{CAB} & \bm{CB} & \bm{D} &...& 0\\
\vdots & \vdots &  &  \ddots &\vdots \\
\bm{CA}^{\frac{L}{2}-2}\bm{B} & \bm{CA}^{\frac{L}{2}-3}\bm{B} &  \dots  & \bm{CB} & \bm{D}
\end{bmatrix}\label{eq:Teoplitz}
\end{equation} 
\end{small}\noindent
and $\bm{X}$
%=[\bm{x}_k, \bm{x}_{k+1}, \bm{x}_{k+2},\dots,\bm{x}_{k+T-L}]^T$
represents the sequence of the state vector with the length of $T-L+1$. 

Then different subspace identification methods can be applied based on (\ref{eq:datamatrixSubspace1})-(\ref{eq:datamatrixSubspace2}), which typically involve two steps: (a) identification of $\bm{\mathcal{O}}_k$ and $\bm{\mathcal{H}}_k$; (b) estimation of system parameter matrices (i.e., $\bm{A},\bm{B},\bm{C}$) from the identified $\bm{\mathcal{O}}_k$ and $\bm{\mathcal{H}}_k$ \cite{Jamaludin2013}. Detailed procedures and comprehensive comparisons of the subspace methods can be found in \cite{katayama2005subspace}.
Besides, recursive stochastic subspace methods are also applied in power grid damping mode estimation and control to realize online adaptiveness \cite{sarmadi2013electromechanicalf,zhang2012adaptive,wu2006multivariable}.

%For example, N4SID projects $\bm{Y}_F$ onto $[\bm{Y}_P; \bm{U}_P; \bm{U}_F]$ and conducts SVD on the part corresponding to the past data. 

Among the above-mentioned subspace methods, CVA \cite{Ghasemi2006, WControl_SSIM_CVA}, N4SID \cite{Ghasemi2006, Jamaludin2013, WideAreaControl_N4SID,Mansouri2018,Yohanandhan2016,Nieto2015,Zhu2016}, and MOESP \cite{YE2013509} may have the bias problem in nature since they work under the open-loop assumption (i.e., control inputs $\bm{u}$ is not correlated to the process noise $\bm{\delta}$ and the observation noise $\bm{e}$). In contrast, the OKID (observer Kalman filter identification) and the variants are free of the bias problem even in the closed-loop condition that the control inputs $\bm{u}$ correlate $\bm{\delta}$ and $\bm{e}$. %\cite{QIN20061502}.  
We refer readers to %\color{black} 
\cite{katayama2005subspace} %\color{black} 
for a detailed review of subspace methods, where the general procedure (including pre-estimation, regression or projection, model reduction and parameter estimation) is summarized.  %\color{red} you referred to other references in the last paragraph. Maybe only one is needed? \color{black}
%\color{black} 
The above-mentioned classical identification is usually conducted with ambient data as the linearization is based on the assumption of small signals in ambient conditions. The efficacy of the identification may degrade during transients.\color{black}
%\color{red}
%Once the state-space model is identified, XX (control method) was applied in []  for (what application)....
%\color{black}

Once the state-space model is identified, mature linear control techniques can be applied. For example, different state-feedback control methods, such as linear quadratic regulator (LQR) and root locus-based control design, can be applied \cite{WideAreaControl_N4SID,6082169,Nieto2015} for wide-area damping control. Model predictive control \cite{YE2013509},   root locus control design\cite{Zhu2016}, residual-based control \cite{zhang2012adaptive}, and PID control \color{black}\cite{Yohanandhan2016} were also designed for damping control based on the identified models. The control parameters can be designed with heuristic optimization such as particle swarm optimization \cite{Yohanandhan2016}.

\subsubsection{Linear System Identification Based on Input-Output Models}

\begin{figure}
\centering
  \includegraphics[width=0.75\linewidth]{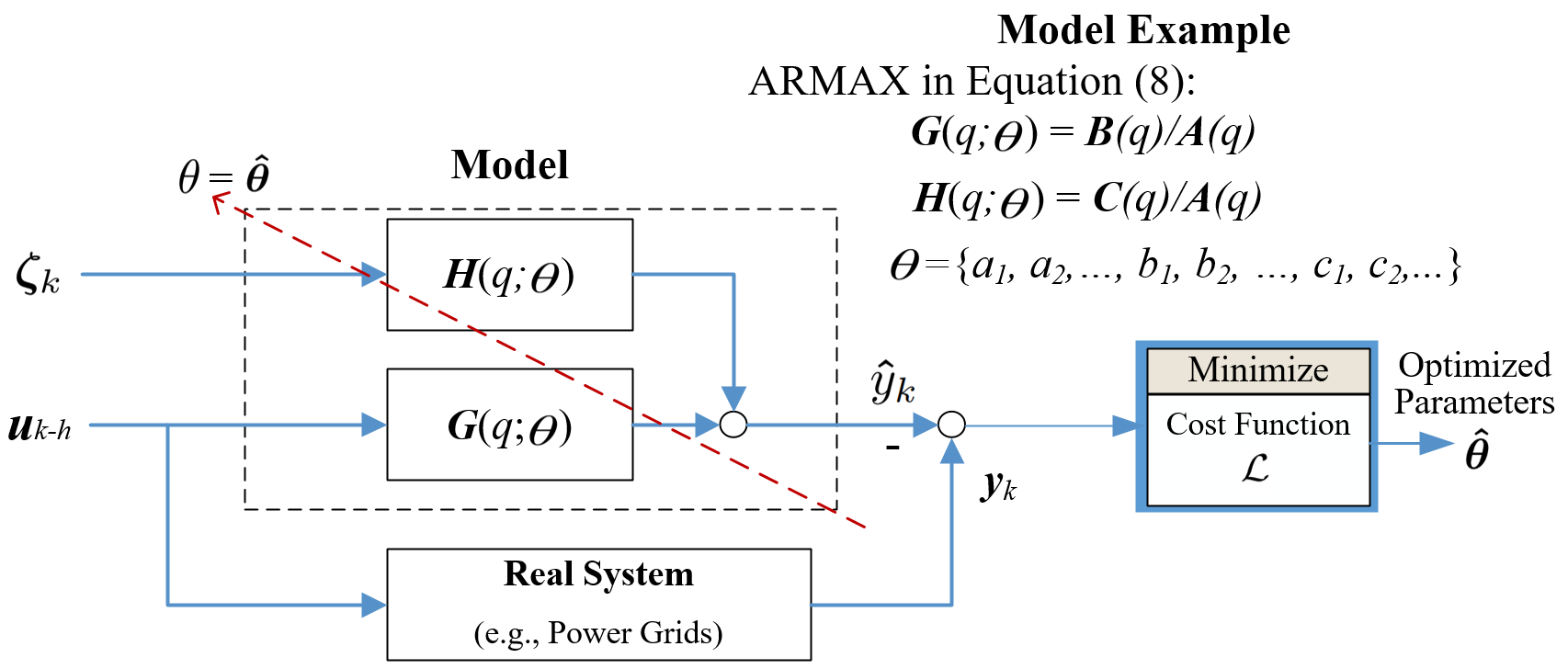}\\
  \caption{The schematic structure of prediction error methods.}
  \label{fig:PEM}
\end{figure}

We consider another special form of the model representation, i.e., the transfer function-based model, which directly describes the time series of input $\bm{u}$ and output $\bm{y}$  as a ``black-box" without the intermediate state $\bm{x}$ in (\ref{eq:id_general}) and (\ref{eq:ctrl}). An example is the autoregressive moving average with external inputs (ARMAX) model:
\begin{equation}
%{A}(q)\bm{y}_{k+h}={B}(q)\bm{u}_k + \mathcal{M}_{ML}(\bm{\chi})+\bm{e}_k
{A}(q)\bm{y}_{k}={B}(q)\bm{u}_{k-h} +{C}(q)\bm{\zeta}_{k}
\label{eq:ARMAX}
\end{equation} \noindent
$A(q) = 1 + a_1 q^{-1} + a_2 q^{-2} + \dots+ a_{n_a} q^{-n_a}$ , $B(q) = b_0 + b_1 q^{-1} + b_2 q^{-2} + \dots + b_{n-1} q^{-n+1}$ and $C(q) = 1+ c_1 q^{-1} + c_2 q^{-2} + \dots + c_{n_c} q^{-n_c}$ are the ARMAX polynomials; $q$ is the delay operator; the parameters $n_a$, $n_b$, and $n_c$ are the orders of the ARMAX model. $h$ is the time delay of the system and $1 \leq h < n_b$. The term $\bm{\zeta}_k$ denotes the white-noise disturbance.  Other types of transfer function-based models, such as output-error models and Box-Jenkin models, are illustrated in \cite{ljung1999system}. These models can be identified with prediction error methods, which aim to find the system parameters that minimize the prediction error \cite{ZHAO2012180, ljung2002prediction}. Fig. \ref{fig:PEM} shows a schematic structure of the prediction error method for a general input-output model, which tunes the model parameter $\theta$ to minimize the loss function of prediction error $\mathcal{L}$.  Generally, the cost function can be defined as $\mathcal{L} = f_c(Q_{cov}(e_{p,k}))$, where $e_{p,k}=y_k-\hat{y}_k$ represents the prediction error;  $f_{c}$ is a scalar monotonically increasing function; $Q_{cov}=\frac{1}{N_{sample}}\Sigma_{k=1}^{N_{sample}}(e_{p,k} e_{p,k}^T)$ 
%denotes the sample variance for single output and 
denotes the sample covariance matrix.
%for multiple outputs. 
%\color{red} check the last sentence \color{black} %\color{red} I think in 1) and 3) we briefly explain how method works. We may also need to briefly explain how the identification works here. \color{black}

%denotes the unmodeled dynamics and is a higher-order nonlinear function of the input $u$ and output $y$. 
%\color{black} As shown in \cite{Liu2017}, the ARMAX is the equivalent discrete transfer function model whose input-output relationship can also be equivalently derived from and the state space model (\ref{eq:process})-(\ref{eq:ssmodel}). \color{black}(TBD)\color{red} are they equivalent or equivalent under some conditions? \color{black}

For power system applications, the authors in \cite{Liu2017} proposed an ARMAX-based damping control. The ARMAX was used to capture dominant inter-area modes of power systems identified in a least-squares fashion. Likewise, the authors in \cite{8882329} assumed that the LVDC (low voltage direct current) system was linear and can be interpreted as a linear difference system equivalent to (\ref{eq:ARMAX}), which was identified with singular value decomposition.
%with the stability-oriented regularization based on the linear matrix inequality technique. 
%\subsubsection{{Linear Identification Based on Transfer Functions}} \color{red} I am not sure whether to combine with the last subsection or not. It seems they belong to ``a general family of model structures'' in the book ``System identification: theory for the user: Lennart Ljung; Prentice-Hall, Englewood Cliffs, NJ, 1999, ISBN 0-13-656695-2''\color{black}
%As discussed in \cite{Liu2017}, the ARMAX model is also equivalent to the transfer function model. 
Another recently developed transfer function-based identification method was proposed in \cite{KARIMI2017227} based on the frequency response (FR) of multivariate continuous-time systems and convex optimization. %\cite{KARIMI2017227}. 
This model-free method was applied to control battery energy storage systems in islanded microgrids to reduce voltage and frequency fluctuations \cite{Ryan2021}. The transfer function was identified in the spectral-analysis method \cite{Barkley2009}, which applied the discrete Fourier transform to the auto-correlations of inputs and output data. %(\color{red} check the previous sentence, too many ``which'' (done)\color{black}) %purely free of the dynamic model. \color{black}
The authors in \cite{8973540} \color{black} modeled \color{black} the power system with continuous-time transfer functions too, which were identified online with  ring-down  data using a two-stage least-squares algorithm to realize the highest regression accuracy indices in both time domain and frequency domain. 

Once the transfer function model %\color{red} does this paragraph correspond to the MIMO model (9)? In that case, will MIMO be better than transfer function? (done) \color{black} 
is identified, classical control techniques, such as output-feedback control with lead-lag compensator \cite{8973540} and networked predictive control \cite{yao2014wide},  are applied for wide-area damping control. Generally, there is a trade-off between identification performance and robustness; a robust controller design is often favorable. For example, $H_{\infty}$ robust control has been applied in microgrid primary control \cite{Ryan2021}. A configuration information-free controller design for linear difference systems was also applied for active load stabilization in \cite{8882329} based on a Lyapunov stability-oriented quadratic function and linear matrix inequality techniques. Similar LMI-based control can be found in \cite{Madani2021}, where some other goals such as power sharing and frequency/voltage restoration were incorporated in the objective function jointly with the stability requirements. In \cite{YangDW2020}, an ARMAX model was identified whereby the estimate of inertia was obtained and sent to the local-area model predictive controllers to realize frequency regulation\color{black}.
Similar to \emph{1)}, the input-output model is linear. Therefore, the identification is suitable for small signals in ambient conditions and usually conducted with ambient data. The efficacy of the identification may degrade during transients. %\color{red} may need to change a place (done). \color{black}
%\color{black}Similar to \emph{1)}, the input-output model is also linear. Therefore, the identification is suitable for small signals in ambient conditions and usually conducted with ambient data. The efficacy of the identification may degrade during transients. \color{red} may need to change a place. 
\color{black}

%\cite{Hou2011} (TBD) 
%\color{black}(TBD, nonlinear/linear, direct/indirect category?)   
Another formulation of multi-input multi-output (MIMO) model is proposed in \cite{Hou2011} based on a novel dynamic linearization technique. Specifically, the system of interest is adaptively linearized as:
\begin{equation}
%{A}(q)\bm{y}_{k+h}={B}(q)\bm{u}_k + \mathcal{M}_{ML}(\bm{\chi})+\bm{e}_k
\Delta\bm{y}_{k+1} =  \bm{G}_k \Delta \bm{u}_k
\label{eq:MFAC}
\end{equation} \noindent
\begin{equation}
%{A}(q)\bm{y}_{k+h}={B}(q)\bm{u}_k + \mathcal{M}_{ML}(\bm{\chi})+\bm{e}_k
\mbox{with } \bm{G}_k = \begin{bmatrix}
G_{11,k} & G_{12,k} & ...& G_{1p,k}\\
G_{21,k} & G_{22,k} & ...& G_{2p,k}\\
\vdots & \vdots & \vdots & \vdots\\
G_{p1,k} & G_{p2,k} & ...& G_{pp,k}\\
\end{bmatrix}, \quad \| \bm{G}_k\| \leq b
\label{eq:MFAC_G}
\end{equation} \noindent
where $b$ is a positive constant; $\bm{G}_k$ is the matrix of pseudo-partial derivative, which can be identified with time-varying parameter estimation algorithms such as modified projection \cite{Hou2011},  least squares with a forgetting factor \cite{Paleologu2008}, and the leakage recursive least-squares method \cite{rogers1996adaptive}. Such a model has been used in the identification for microgrid primary control realizing a robust local voltage control that is not sensitive to the variations of the system parameter, structure, and control delay \cite{zhang2015data}. The model has also been used for power system stabilizers in \cite{lu2015wide} with improved dynamic performance for damping low-frequency oscillations under various operating conditions. Because the linearization is adaptive and the model can be identified in a rolling-based fashion, it can also handle nonlinearity and can be conducted for both ambient and transient data. However, the effectiveness based on transient data relies on the window length of data and the incurred inertia, which may hinder a timely model update.

\subsubsection{Linear Identification Based on Ornstein-Uhlenbeck Process}
%\color{red} this subsection can be the last subsection before direct linear identification and control \color{black}

The aforementioned identification techniques estimate the system parameters with minimized error in a mathematical sense while the mathematical solution is not unique. Therefore, the identified models may not reflect the real system state variables and the estimated parameters are not guaranteed to be the parameters of the true physical system, making the control
design challenging. %Besides, the estimated parameters are not guaranteed to be the parameters of the true physical system.  \color{red}Comment on what exact control method is applied based on the model identified by what method in previous works (at least a few examples).
Leveraging the statistical properties of multivariate Ornstein-Uhlenbeck (OU) process, the data-driven identification methods  %Incorporating certain physical information into the identification can help can better describe the real state variable evolution. In 
developed in \cite{Georgia2021,Ilias2018,Ilias2020,Guo2021} can estimate the system state matrix of the true physical model and extract modal properties and essential parameters.  %the power system dynamics are modeled with a physics-informed structure and the assumption that the load is perturbed by independent Gaussian noise. %\cite{lange_physical_2006, Ilias2020}. . 
Assuming that the aggregated load powers experience Gaussian stochastic perturbations, the power system dynamic model in ambient conditions can be represented as
%That is
\begin{small}
%\begin{subequations}
\begin{equation}
%(Process \quad  model) 
\bm{\dot{x}} = \bm{A}\bm{x}+\bm{B}_P\bm{\xi}%+\bm{B}_C\bm{u}
%\bm{\dot{x}} = \bm{A}\bm{x}+\bm{B}_P\bm{\xi} %+\bm{\delta}_k
\label{eq:OU}\end{equation}
%\begin{equation}
%\bm{y}_k = \bm{C}\bm{x}_k +\bm{D}\bm{u}_k+\bm{e}_k
%\label{eq:ssmodel}\end{equation}
%\end{subequations}
\end{small}\noindent
%where the state $\bm{x} = [\Delta \bm\delta, \Delta \bm\omega]^T$, with $\Delta\delta$  the vector of generator rotor angle deviations and $\Delta\omega$ the vector of generator rotor speed deviations.
where the states $\bm{x}$ can be %include 
the deviations of generator rotor angle, generator rotor speed in wide-area damping control \cite{Ilias2018,Ilias2020,Guo2021}, or be the deviations of bus voltage magnitude and phase angle in wide-area voltage control \cite{Georgia2021} or load identifications \cite{Pierrou2020}. $\bm{\xi}$ are mutually independent standard Gaussian random variables,  representing the stochastic load variations; $\bm {B}_P$ is the noise intensity matrix. 
%$\bm {B}_P \bm\xi$ are the stochastic state variations that follow Gaussian distributions, with $\bm\xi$ the random variables that follow standard Gaussian, and $\bm {B}_P$ the coefficient matrix. 
Thus, the state $\bm{x}$ in (\ref{eq:OU}) follows a multivariate OU process. The parameter $\bm A$ corresponds to the Jacobian matrix of the true physical model containing significant information like all oscillation modes, mode shapes, participation factors, inertia and damping constants, dynamic load time constants, network topology parameters, etc. %is the state transition matrix with the physical meaning characterized by Jacobian, inertia coefficients, damping coefficients and dynamic load time constants, etc. 
%$\bm B_C$ is the control matrix which is assumed known.
%Before the control is enabled (i.e., $\bm{u}=0$),  

The system parameter $\bm A$ can be analytically obtained by applying the regression theorem of OU process, which reflects the dynamical evolving of state covariances and relates the statistical properties of state variables with $\bm A$.  %reflects the dynamical evolving of state covariances. The parameter matrix $\bm A$ can be estimated based on the Regression Theorem with respect to the 
%$\tau$-lag state covariance matrix $\bm{G}(\tau)$. That is
Specifically, 
\begin{small}
%\begin{subequations}
\begin{equation}
\frac{d}{dt}\bm{G}(\tau) = \bm{A}\bm{G}(\tau),\mbox{with } \bm{G}(\tau)= <[\bm{x}(t+\tau)-\bar{\bm{x}}][\bm{x}(t)-\bar{\bm{x}}]^T>\\
\label{eq:regression}\end{equation}
%\end{subequations}
\end{small}\noindent
%Based on the matrix 
Hence, $\bm{A}$ can be obtained uniquely and analytically by solving (\ref{eq:regression}) according to \cite{Sheng2020}:
\begin{small}
%\begin{subequations}
\begin{equation}
\bm{A}=\frac{1}{\tau}log[\bm{G}(\tau)\bm{G}(0)^{-1}]
\label{eq:rtA}\end{equation}
%\end{subequations}
\end{small}\noindent
%The identification of $\bm{A}$ as in 
Equation (\ref{eq:rtA}) tactfully provides a way to identify the true physical model $\bm A$ from the statistical properties of measurement data.  %uses the statistical properties of measurement data with the physical properties of the dynamic power system model. 
Alternatively, $\bm{A}$ can be obtained based on the Lyapunov function that the stationary covariance matrix $\bm{G}(0)$ of multivariate OU process satisfies \cite{Ilias2020}:
\begin{small}
%\begin{subequations}
\begin{equation}
\bm{A}\bm{G(0)} + \bm{G(0)}\bm{A}^T = -\bm{B}\bm{B}^T
\label{eq:lyp}\end{equation}
%\end{subequations}
\end{small}\noindent
%where $\bm{C}_{xx}$ %is the covariance matrix 
%can also be estimated with data.
if $\bm{B}$ is known. It was shown that the true system parameters were approximately obtained by using both identification methods (according to (\ref{eq:rtA}) and (\ref{eq:lyp}) respectively); then the identified model can be directly utilized in control design.
%[Ornstein–Uhlenbeck process-based methods \cite{Georgia2021,Ilias2018,Ilias2020,Guo2021}]
%[Frequency Response-Based Data-Driven Control\cite{KARIMI2017227,Ryan2021}]: 
Note that the OU-based methods are conducted with ambient data to identify the small-signal model of interest. Different from \emph{3.1.1)} and \emph{3.1.2)}, this type of method can help obtain a unique identification solution that corresponds to the true physical small-signal model. %parameters. 
\color{black}
Even though the model structure has certain physical interpretation,  %and accurate physical parameter (such as damping and inertia) estimation may be uniquely obtained in the identification process, 
the methods are in nature ``data-driven". %and ''(first principle/physical) model-free".
\color{black}

Once the system of interest is identified by the methods stated above, conventional linear optimal or robust control methods can be applied. For example, modal linear quadratic regulators \cite{Guo2021} and 
pole placement \cite{Ilias2018,Ilias2020} have been applied in wide-area damping control.
%Linear quadratic Gaussian (LQG) has been applied in \cite{}.
%Pole placement has been applied in wide-area damping control \cite{Ilias2018,Ilias2020}
Because the OU-based methods can identify the true system model, the wide-area damping controls designed based on the identified model can achieve full decoupling of modes and damp all critical modes simultaneously without affecting the others. Optimal control has also been applied in wide-area voltage control \cite{Georgia2021}.

%Generally, there is a trade-off between identification performance and robustness; a robust controller design is often favorable. For example, $H_{\infty}$ robust control has been applied in microgrid primary control \cite{Ryan2021}.

%\color{red} may think about how to group this method in Figure 5? (added as a new branch in Fig. 5) 
\color{black}
\subsubsection{Loewner Method}
Most of linear models and identification are based on a representation that can describe the system dynamics in the whole frequency range (e.g., the state space representation or transfer function input-output models). 
For a selective-frequency range, Loewner interpolation was commonly used in black-box modeling of large MIMO microwave structures that show the capability for rational interpolation of frequency data \cite{kassis2016passive, kabir2012macromodeling}. In 2015, a tangential interpolation framework based on Loewner matrix pencil was first used in power systems for modeling frequency-dependent network equivalents that can describe electromagnetic transients (EMT) \cite{gurrala2015loewner,Dmitry2020fr}. In 2019, the Loewner interpolation method was used for the reduced-order model and identification for electro-mechanical models of multi-machine power systems \cite{rergis2018loewner}. The method can realize more accurate identification in a particular frequency range than other techniques. However, the Loewner methods are only suitable for linear time-invariant systems (i.e., assuming small signals at specific operating conditions), which may not hold for modern low-inertia power grids with increasing volatile renewables. Likewise, the identification based on Loewner method for a selective-frequency range was compared with the conventional eigensystem realization algorithm for a New-England 10-machine test system in 2020. The results show the performance enhancement in terms of the modal shape and the absolute error of identified frequency responses \cite{Zelaya2020}. After applying Loewner methods, the identified transfer functions for different frequency ranges corresponding to different potential operating conditions then can be applied for robust controller design.
\color{black}

\subsection{Direct Linear Identification and Control} Although the linear identification and control tasks are usually conducted sequentially as aforementioned, effort has been made to realize the two tasks simultaneously \cite{Dorfler2022} to benefit the global optimality of joint identification and control. For example, data-enabled predictive control (DeePC) with quadratic regularization is investigated for local converter control \cite{HUANG2021192}. It identifies the system parameters and designs control signals together by solving a single optimization problem of DeePC. In summary, the DeePC for the state space model (\ref{eq:process})-(\ref{eq:ssmodel}) can be written in a generic form\cite{Coulson2019,HUANG2021192}:
\begin{small}
\begin{subequations}
\begin{equation}
\min_{\bm{g},{\sigma}_y,\bm{u},\bm{y}} \Sigma_{k=0}^{N-1}\|\bm{u}\|^2_R + \|\bm{y}-\bm{y}_r\|_Q^2 + \lambda_y \| {\sigma}_y\|_2^2
\label{eq:DeePC}\end{equation}
\begin{equation}
 %\bm{y}_k = \bm{C}\bm{z}_k+\bm{e}_k
 \mbox{subject to } [\bm{U_P}, \bm{Y}_P, \bm{U}_{F}, \bm{Y}_{F}]^Tg = [\bm{u}_{ini},\bm{y}_{ini}+\sigma_y,\bm{u},\bm{y}]^T
\label{eq:DeePC2}
\end{equation}
\begin{equation}
\bm{u} \in \mathcal{U}, \quad
\hat{\bm{y}} \in \mathcal{Y}
\label{eq:DeePC3}\end{equation}\end{subequations}\end{small}\noindent
where the operator $\|a\|_X^2$ denotes the quadratic form $a^TXa$. $\lambda_y$ is the regularization parameter. ${\sigma}_y$ is an auxiliary slack variable to guarantee the feasibility of solving the optimization problem. $N$ is the prediction horizon. $\bm{u}_{ini}$ and $\bm{y}_{ini}$ are the most recent input and output trajectories of length $T_{ini}$. 
%$\bm{U}_P$, $\bm{Y}_P$, $\bm{U}_F$ and $\bm{Y}_F$ represent the "past" input and output data, as well as the "future" input and output data, respectively. $[\bm{U}_P,\bm{U}_F]^T$ constitutes the Hankel matrix of the input $\bm{u}$, and $[\bm{Y}_P,\bm{Y}_F]^T$ constitutes the Hankel matrix of the output $\bm{y}$. For example, the Hankel matrices with the depth of $T$ are as follows:

The DeePC can be seen as a special form of (\ref{eq:id_ctrl})-(\ref{eq:uy_constraint_direct}), with $\bm{g}$ corresponding to the parameter $\theta$. The DeePC solves the convex optimization problem (\ref{eq:DeePC})-(\ref{eq:DeePC3}) in a receding horizon manner with a finite number of data samples to predict future trajectories. \color{black} See details in \cite{Coulson2019}. The DeePC has been applied in modern power grids, such as local converter control \cite{HUANG2021192}, decentralized damping control \cite{Huang2022}, and frequency regulation \cite{zhao2021data}. The DeePC can also be used to guide the training phase of RL with improved sample efficiency, It was applied to load frequency control and the control performance was improved \cite{zhao2022}. %These applications show the potential to control performance improvement.
%\color{red} any particular application in power systems? \color{black}
\color{black} Another example of direct data-driven control is a multivariable linear parameter varying (LPV) controller  applied in islanded microgrid secondary-primary control, resulting in damped oscillations in frequency and power dynamics \cite{madani2021data}. Different from DeePC, the synthesis process is based on the frequency domain rather than the time domain.\color{black} 

\color{black}
It has been shown in the IEEE task force paper \cite{sanchez2012identification} that linearized models around the operating points can yield an approximation behaving at a reasonable level of accuracy 
%for both transients and normal operation 
using transient data %\color{red} it seems you use ``transient" in section 2 (fixed, also added a note that the transient sometimes termed as ring-down in Section 2) \color{black} 
and ambient data, respectively. %Even for modern low-inertia power grids with higher penetration of renewables, linear models could also work so long as the introduced model uncertainty is reasonably bounded. 
Thus, the linear identification and control methods stated above, either indirect or direct, are still useful and credible in applications like electromechanical modes identification, dynamic load modeling, and wide-area damping control. %(\color{red} give a few more applications\color{black}). 
Also, linear methods are advantageous in two respects: (i) their parameters can be identified optimally with the mature theory foundations of linear systems; (ii) it is easy to realize adaptiveness due to the parameter identification efficiency in linear models. %(iii) linear models are capable enough and intuitive to model oscillatory modes in conventional power grids, and generally black-box linear models tend to have lower model variance than black-box nonlinear models.

%(iv) online identification can adaptively bound unmodeled nonlinear dynamics with the assumption that the dynamics evolving between adjacent time steps can be well approximated by a linearized model.
Nonetheless, %there are still limitations for linear methods 
linear system identification and control may still face challenges in power system applications: %modern power grid applications:
(i) Modern power grids, with the integration of increasing renewable energy sources, exhibit low inertia and fast-changing operating conditions %that are susceptible to 
subject to large disturbances, %As a result, %the system behavior may deviate from linearity, rendering linear models inaccurate. %Modern power grids with increasing renewables are characterized by low inertia and fast-changing operating conditions subject to relatively large disturbance. 
%the system may not behave in a relatively linear fashion thus, 
making the linear models inaccurate. 
(ii) %modern 
The saturation of control signals and the diverse interactions between controllers  may introduce nonlinearity and uncertainty, 
%Introducing nonlinearity through control signals, such as saturation, 
compromising identification accuracy and can even result in ineffective modeling \cite{Smith1993}.
%When control signals used for identification is introducing nonlinearity (e.g., the control signals get saturated), then the identification accuracy is compromised or even leads to ineffective modeling \cite{Smith1993}. %\color{red} Did you discuss saturation? \color{black}

\section{Nonlinear System Identification and Control}
\color{black}
%\textbf{\emph{Nonlinear data-driven control in power systems}}. The nonlinearity interaction of power systems emerges due to the large-signal disturbances, and the nonlinear nature of renewable energy (such as wind and solar), electric vehicles (EVs), and responsive load \cite{Chen2022_RLreviewPS}. Examples of the power system applications include frequency regulation \cite{Younesi2020}, reactive power control \cite{KOU2020114772}, automatic voltage control \cite{Yin2021}, EV charging scheduling \cite{Wan2019,Li2020}, battery management \cite{Bui2020}, residential load control \cite{Chen2021,Ruelens2017}, microgrid secondary dynamics control \cite{Gong2022}, and transient frequency control \cite{KORDA2018297}, cyber-resilient control \cite{Chen2022}, etc.
\color{black}

%Generally speaking, the large-scale integration of renewables and electronic-interfaced DERs increases the modern power grid complexity,  uncertainty and nonlinearity. %nonlinearity and uncertainty.
%\color{black} resulting from the diverse nonlinear energy resources, relatively large disturbances and low inertia compared to conventional power systems. %\color{red} large disturbance?\color{black} \color{black}  
\color{black} To address the aforementioned complexity, uncertainty and nonlinearity of modern power grids, \color{black}%issues, 
nonlinear data-driven control methods gain increasing attention. %and are presented below.
They are categorized as: (A) pure machine learning-based methods; (B) physics-informed machine learning; and (C) Koopman-based methods. Generally, the category (A) uses ``black-box" machine learning to model nonlinear system dynamics and control. The category (B) intends to integrate physical information into the machine learning training or model structure design to realize more reliable modeling. %performance improvement.
\color{black}%\color{black} uses physics-informed techniques with machine learning (e.g. NNs) to make the machine learning more reasonable. 
The category (C) considers the Koopman operators to map nonlinear dynamics to linear state spaces, whereby mature linear control methods can be readily applied. %data-driven modeling and control can be done in a linear fashion. %for more reasonable modeling and control. %\color{red} no offline training? \color{black} 
The three categories are elaborated below. %in what follows. \color{black}
\subsection{Pure Machine Learning-Based Methods}
%\color{black}[Reinforcement Learning]\color{black}[Reinforcement Learning]
\color{black} Generally, machine learning includes reinforcement, supervised and unsupervised learning. 
Reinforcement learning, as a branch of machine learning that can bridge control, becomes popular nowadays for the applications of data-driven control of power grids \cite{Chen2022_RLreviewPS,Zhang2020_RW,She2022}. The learning process generally depends on offline simulators to interact with rather than probing data to real systems to generate training data \cite{Chen2022_RLreviewPS}. However, the RL training is often time-consuming, and the control performance rests on the simulator accuracy and the optimization error associated with highly nonlinear models. 
%and limited time to solve the optimization. 
%\color{red} what is limited solving time? and why is it related to control performance? (modified. it refers to the run time requirement. If the application requires the solving should be done in 10mins, then the RL may not be able to finish the training)\color{black}
%By interacting with the simulation environment to pursue Bellman optimality, RL can yield an optimized control policy that can be deployed. 
%The control optimality is theoretically sound while without performance guarantee due to the existence of simulator uncertainty/modeling error and optimization error \cite{haykin_neural_2010}. 
%\color{red} The previous few sentences are exactly the same as those in Section 2. need to be modified or rephrased at least. (done)\color{black}

Supervised learning can also be used for data-driven modeling for control. However, incorporating control input data in the supervised learning-based model, which is then used for control, is not popular because of data issues associated with control input channels. It is not feasible to probe large testing signals due to the practical operation requirements (especially for safety-critical systems like power grids). Even with low-level test signals, the closed control loops in power grids could trap the state-and-control pairs and make them ill-distributed \cite{nguyen2022globally}, thus lacking rich information for universal learning machines such as NNs to generalize the impact from and to control input channels. In a sense, supervised learning is more often seen in power grid modeling without external control inputs rather than grid control applications. 
%Besides, universal learning machines such as neural networks are not physically interpretable for controller design. 
Details of the RL and supervised learning-based methods are provided in what follows.
%Reinforcement learning, as a branch of machine learning that can bridge control, becomes popular nowadays for the applications of data-driven control of power grids \cite{Chen2022_RLreviewPS,Zhang2020_RW,She2022}. The learning process generally depends on offline simulators rather than probing data to real systems \cite{Chen2022_RLreviewPS}. By interacting with the simulation environment to pursue Bellman optimality, the reinforcement learning can yield optimized control policy that can be deployed. The control optimality is theoretically sound while without performance guarantee due to the existence of simulator uncertainty/modeling error and optimization error \cite{haykin_neural_2010}. 
\color{black}

\textbf{\emph{Reinforcement learning (RL)}}.
RL is a branch of machine learning regarding how to learn control policies by interacting with the environment to explore and exploit inherent model structure according to Bellman optimality \cite{haykin_neural_2010,brunton2022data}. In RL, the system states are assumed to evolve  according to the Markov decision process (MDP), i.e., the probability of the system occurring in the current state is determined only by the previous state \cite{haykin_neural_2010}. Specifically, an MDP consists of a set of states $s \in \mathcal{S}$, actions $a \in \mathcal{A}$, rewards $r \in \mathcal{R}$, the transition probability $Pr(s',s,a) = Prob(s_{k+1}=s'| s_k=s,a_k=a)$, and the policy function $\pi(s,a) = Prob(a_{k+1}=a |s_{k+1}=s)$.  Thus, the state transition under the policy $\pi$ is
\begin{equation}
Pr(s',s,\pi) = \Sigma_{a \in \mathcal{A}}\{\pi(s,a)Pr(s',s,a)\}
\label{eq:MDP}
\end{equation} \noindent

\begin{figure}
\centering
  \includegraphics[width=0.65\linewidth]{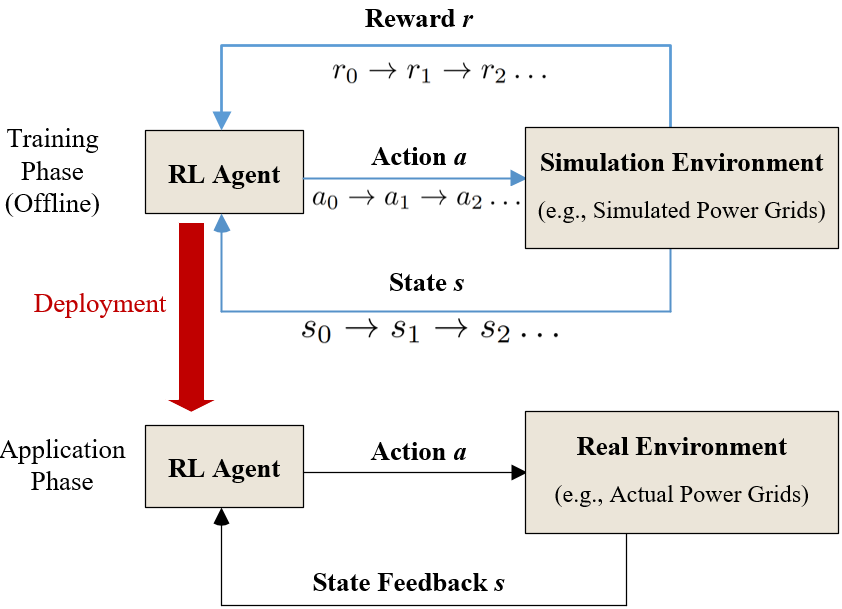}
  \caption{\color{black}The schematic framework of RL.}
  \label{fig:RL}
\end{figure}

The above MDP notation generalizes a Markov process to incorporate actions and rewards for RL, which can be used to describe any nonlinear system decision-making/control of interest. \color{black} Fig. \ref{fig:RL}  shows the generic scheme of RL agent that learns the system through interaction with the environment (i.e., offline simulators), and then the learned agents are deployed to the real environment for online operation. \color{black} 
Note that the RL agent can be either based on universal surrogate models (e.g., actor-critic with deep NNs, approximate dynamic programming, deep Q-learning, etc.), or uses purely model-free paradigms (e.g., Q-learning, SARSA, policy gradient, etc) \cite{brunton2022data,haykin_neural_2010}.

In practical applications of power grids, Q-learning is among the most popular RL techniques \cite{Younesi2020,Chen2021,Yin2021,Silva2020}. However, Q-learning is not inherently suitable for continuous action spaces \cite{brunton2022data}. In addition, the action and reward functions for RL are often unknown or of high uncertainty. Therefore, actor-critic methods gained attention \cite{Chen2022, Khooban2021} and are favorable because of the iterative exploration and exploitation of the ``actor" and ``critic". That being said, the action and reward functions are approximated with two separate learning machines (e.g., various NNs, and support vector machines), which are trained with abundant data. In a broader sense, such ``actor-critic" can be seen as a direct data-driven control paradigm equivalent to (\ref{eq:id_ctrl})-(\ref{eq:uy_constraint_direct}) that identifies the system and generate control actions directly. However, the universal learning machines under the general RL framework often suffer from training efficacy and efficiency problems due to the unconstrained or loosely-constrained learning space: (a) the trained models may not well describe the underlying physical process well due to modeling error and training error; (b) the computational cost of training is high. 
\color{black} The former issue can be mitigated by physics-informed techniques and Koopman-based methods that will be discussed in Sections 4.2-4.3. While the latter can be naturally lifted by parallel computing when applicable. For example, a wide variety of scenarios in power grids can be created in simulation to train DRL.
For example, the authors in \cite{Huang2020} parallelized training tasks based on environment study cases (e.g., fault scenarios in power grids). Furthermore, they proposed a two-layer parallel scheme \cite{HuangR2022} that supports task parallelism based on environment and learning parallelism based on a parallel augment random search algorithm, which was successfully applied in derivative-free DRL for emergency voltage control. It can realize well-structured and more effective parameter exploration (larger learning rates and fewer hyper-parameters to tune) than conventional action-space exploration in DRL, improving the computational scalability and accelerating the training (e.g., training time reduced by 136 times on the IEEE 300-bus system). Likewise,  a parallel framework that employs multiple workers simultaneously interacting with power grid simulators was adopted for autonomous voltage control in \cite{Xu2020}, which shows improved training efficiency and stability.
%\color{red} (DRL with parallelism was added) \color{black}
%The parallel computing can facilitate the multi-task training of deep reinforcement learning. A wide variety of scenarios in power grids can be created in simulation to train DRL. After being trained, DRL possesses certain level of generalization capacity and can be used for some new scenarios. However, such adaptiveness is limited as it highly depends on the simulator data quality, and the system operating conditions and topology may change over time and may be not known by system operators. The old simulation data may not be able to reflect the underlying change of physical systems; thus the fast adaptation to new operating conditions is compromised.

\color{black}As RL is a branch of mature machine learning techniques which have been well illustrated in many other works in the field of power systems,  such as decentralized resilient secondary control \cite{Chen2022,Younesi2020}, microgrid frequency regulation\cite{Younesi2020,Khooban2021}, microgrid power dispatch \cite{Hao2021}, reactive power control \cite{KOU2020114772}, battery management \cite{Wan2019,Li2020,Bui2020}, multi-area AGC \cite{Singh2017, Yan2020_FR}, and wide-area damping control \cite{Yousefian2018}, \color{black}
%\cite{KOU2020114772, Younesi2020, Khooban2021, Hao2021, Singh2017, Yan2020_FR, Bui2020, Wan2019, Li2020, Chen2022, Yousefian2018}, 
they will not be detailed in this paper. Interested readers are referred to \cite{Chen2022_RLreviewPS, Zhang2020_RW, She2022} for comprehensive reviews of RL in power system/microgrid applications. %and their applications in power grids.
%\color{red} shall we follow what we did in previous sections to comment on power system applications briefly?
\color{black}

%[Imitation Learning] (TBD)
%Another category of machine learning-based control rests with imitation learning (IL) \cite{IL2017_ACM}, which has a key difference from RL: i.e., the supervised learning is used in IL to directly learn the mapping between the observations and desired actions. That being said, the model structure is fixed with a learning machine which is then trained to imitate the correct actions of the underlying system with the feedback data of states and actions.

%[Machine Learning to Model Residual Dynamics]
%As RL and IL are mature machine learning techniques that have been well illustrated in many other works, they will not be detailed in this paper. Interested readers are referred to \cite{} for comprehensive introductions of the RL and IL methods. %and their applications in power grids.

%\subsubsection{Online/Offline Framework}

%$a_0 \rightarrow a_1 \rightarrow a_2 \dots$
%$s_0 \rightarrow s_1 \rightarrow s_2 \dots$
%$r_0 \rightarrow r_1 \rightarrow r_2 \dots$

\begin{figure}
\centering
  \includegraphics[width=0.65\linewidth]{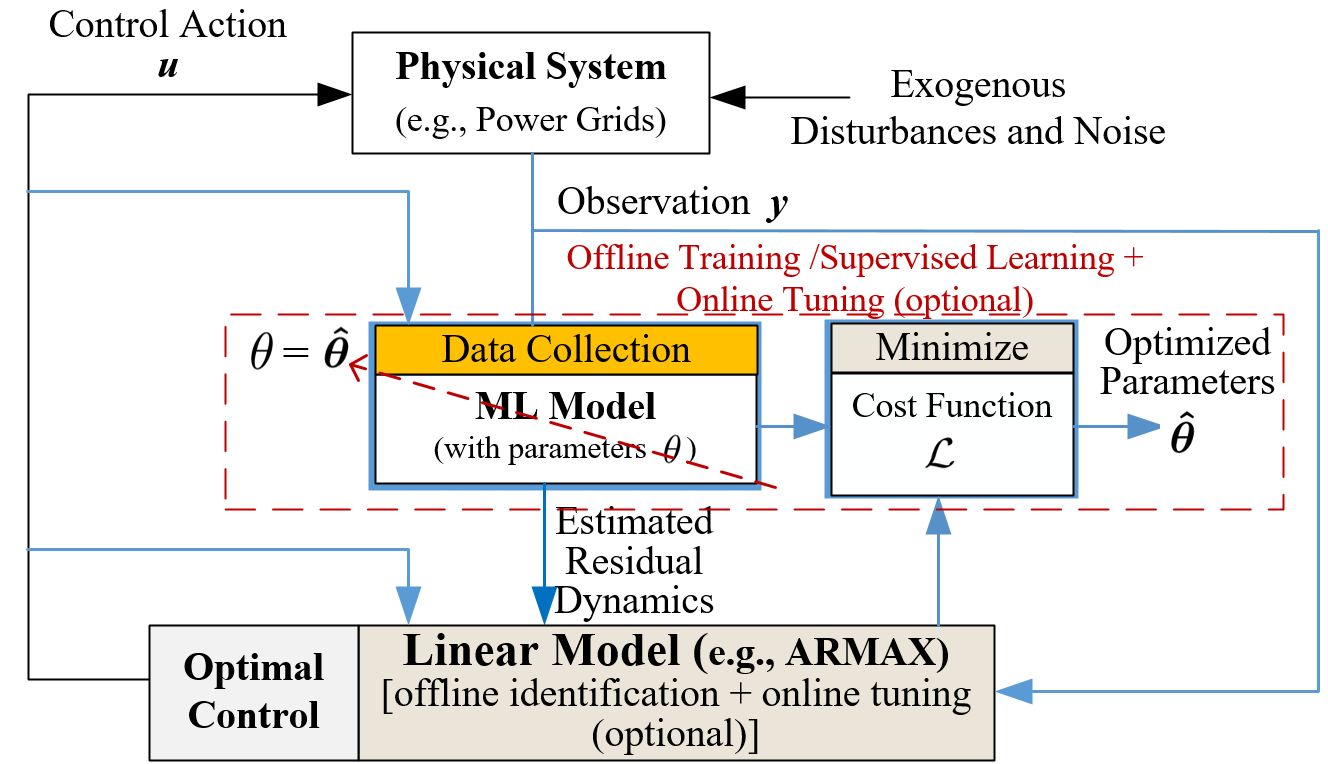}\\
  \caption{The schematic framework of \color{black} supervised learning based modeling for control \color{black} with a nonlinear learning machine and a linear model.}%\color{red}What is the output of the learning machine (one of the input to physical system)?\color{black} TBD}  %explain residual dynamics in the main context
  \label{fig:SL}
\end{figure}

\textbf{\emph{Supervised Learning-Based Method}}. Universal machine learning models, such as neural networks (NNs), can be used in power grids to model nonlinear dynamics \cite{Zhang2010}. 
%\color{black} The word "supervised" means a learning paradigm for problems where the available data consists of labeled examples whereby the learning machine can be trained under the supervision of the label.\color{black}  
For example, the authors in \cite{Ma2021} used a linear dynamic model ARMAX to describe the microgrid voltage dynamics first and then trained NNs to compensate for the residual nonlinear dynamics on top of the linear model, whereby optimal control is applied to realize stabler and faster microgrid voltage control. \color{black} Fig. \ref{fig:SL} shows the schematic framework of supervised learning-based modeling for control with pure ML, where the learning machine is trained offline based on a user-defined cost function $\mathcal{L}$. Note that the linear model (e.g., ARMAX) may keep updating by online identification to realize fast adaptiveness. The ML model parameters may also be finely tuned online around the offline-trained model to realize a certain level of adaptiveness. \color{black} This framework can be seen as a special form of the sequential data-driven control (\ref{eq:id_general})-(\ref{eq:uy_constraint_c}). It has been shown that the well-trained machine learning model can capture power system nonlinearity with supervised learning given abundant data \cite{Ma2021,kazemlou2014decentralized}. Besides the applications in power grids, the NN-based control is also used to model and control residual nonlinear dynamics of drone landing \cite{Shi2019}, where spectral normalization is used to stabilize NN training and to improve generalization capacity. Conventional loss functions for machine learning include root-mean-squared error, mean absolute error, cross-entropy, KL-divergence, etc, which can be used for different training algorithms such as Levenberg-Marquardt (LM) algorithm, stochastic gradient descent (with momentum) (SGD or SGDM), and adaptive moment estimation (ADAM) \cite{haykin_neural_2010,brunton2022data}.

\begin{figure}
\centering
  \includegraphics[width=0.65\linewidth]{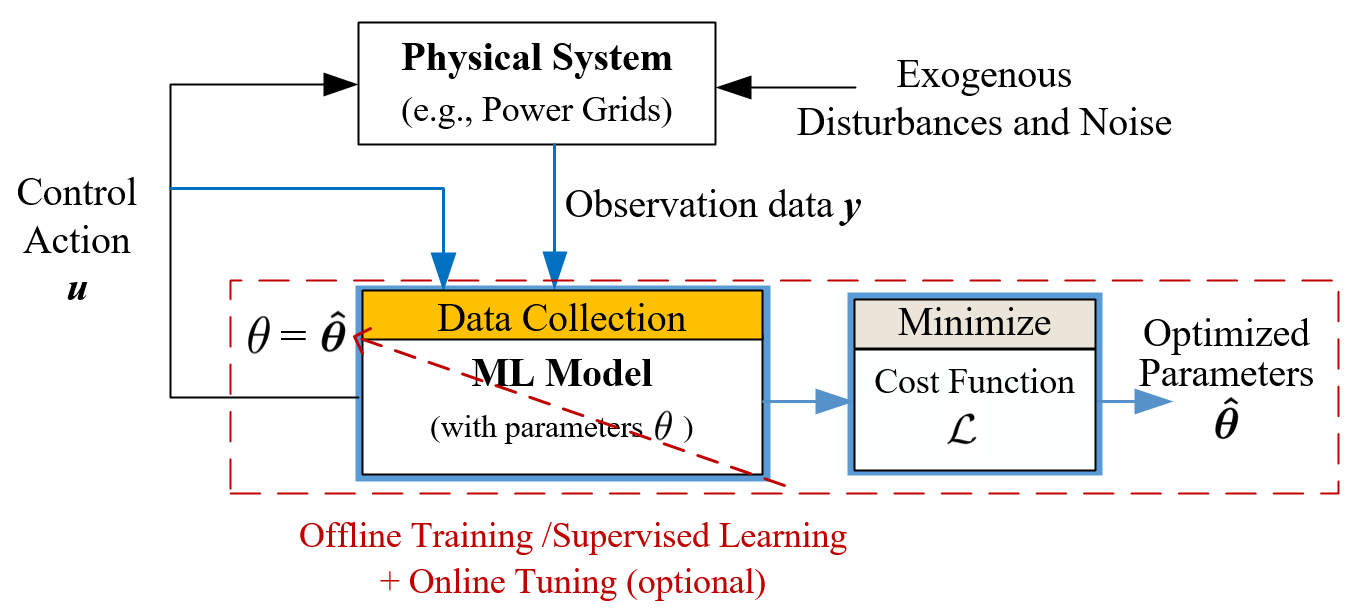}\\
  \caption{The schematic framework of \color{black} supervised learning-based modeling for control \color{black} with nonlinear learning machines only.}
  %\color{red}What is the output of the learning machine (one of the input to physical system)?\color{black}} TBD
  \label{fig:SL2}
\end{figure}

\color{black} The learning machines can also be used to model the uncertainty of systems dynamics and then be directly used to regulate control signals as shown in Fig. \ref{fig:SL2}. The offline-trained ML models may also be finely tuned with new online data to realize certain level of adaptiveness. 
Such an idea of offline pre-training and online-tuning matches the concept of transfer learning in machine learning community, which is a learning paradigm aiming to transfer the knowledge generalized from the data of existing systems to the learning of new similar systems by offline training and the learning machine parameters are fine-tuned when new data come \cite{yosinski2014transferable}. 
\color{black} This paradigm has been used in \cite{Zheng2022,kazemlou2014decentralized,shamsollahi1999application},
%\color{red} how about \cite{kazemlou2014decentralized}?
where online-tuned NNs are used for identification and control to realize accurate power sharing and enhanced stability, respectively. 
%\color{black} Similarly, a NN with online learning is used to learn decentralized control policies to enhance stability of interconnected dc distribution system \cite{kazemlou2014decentralized}. \color{black} 
The online tuning of NNs in \cite{Zheng2022,kazemlou2014decentralized} can also be seen as a special form of the direct data-driven control (\ref{eq:id_ctrl})-(\ref{eq:uy_constraint_direct}), as the NNs are trained to model to obtain optimized parameters and to generate \color{black} control signals simultaneously. NN-based control has also been used in a standalone DC microgrid where the NNs are trained offline to directly output desired control signals for individual converters of distributed energy resources to enhance microgrid voltage stability \cite{dong2018artificial}. The optimal Koopman operators and the identification need further investigation. \color{red} \color{black}

Pure machine learning methods can be conducted with both ambient and transient data. However, the data should be informative enough to guarantee the generalization of learning machines. \color{black}Besides, pure machine learning methods resolve the control policies as ``black-boxes" with no explicit physical interpretation. Therefore, they are often analytically and computationally less tractable and less practical for real-time and safety-critical scenarios in power grids. Some physics-informed learning methods have been proposed to address the issues, which are presented as follows.

\subsection{Physics-Informed Machine Learning Methods}

\color{black}
%Although general learning machines such as neural networks are of full capacity to fit any nonlinearity, there are a few challenges that remain and hamper their applications in power systems:
%\emph{1) The data volume for training is usually high, the model is relatively big and the training time is therefore long. In a sense, they are more suitable for relatively unchanged environment with fixed topology and predefined disturbances \cite{She2022} to have adequate data and conduct timely training}.
%\emph{2) The data may not contain adequate information to be learned. The probing signals are hard to design, as it may not be sufficient enough to fully emerge inherent nonlinear characteristics due to noisy environment and sequential dependence.  Low-level probing signals are favorable in power grids in consideration of power quality.}
%\emph{3) Offline training of the learning machines may provide optimal or local-optimal parameters in terms of data fitting accuracy in a mathematical sense, however, the physical interpretablility or explainability lacks and thus compromises the model fidelity and controller design.}
%\emph{4) In certain scenarios, obtaining a ``warm-up" offline-trained model to start with is problematic due to the lack of a large volume of on-field data and the run time requirements.}
%Therefore, they are often analytically and computationally less tractable and less practical for real-time and safety-critical scenarios in power grids. \color{black}

\color{black} The physics-informed machine learning can inform the learning machine of certain physics either through the training, the model structure design, or both. Thus, they can enhance the interpretability or explainability with more physics-consistent solutions.
In what follows, we discuss physics-informed methods from the two perspectives. Interested readers are also referred to \cite{Huang2022RW} for more detailed information regarding physics-informed NNs that have been used in power grids.\color{black}

\subsubsection{Physics-Informed Training}
%Conventional subspace identification methods \cite{QIN20061502} can be applied directly in the Koopman state space. This is so because, the Koopman-enabled linear time-invariant (LTI) systems are characterized as shift-invariant subspaces within an ambient space of time series. The identification is to find such a low-dimensional feature from data. For example, subspace is to extract parametric models from the range and null spaces of a low-rank data Hankel matrix \cite{Dorfler2022,QIN20061502}.
In the training phase of machine learning, different regularization and relaxation techniques\cite{Dorfler2022,Gupta2022} can be incorporated to avoid over-fitting and increase physics interpretability. For example, different norms can be added in the regularization terms such that the system model complexity can be constrained under certain mathematical and physical interpretations (e.g., stability or other specific physical constraints). In general, the identification loss function with the regularization terms can be written as
\begin{small}
%\begin{subequations}
\begin{equation}
\begin{aligned}
\mathcal{L}_{id} = &\mathcal L(\bm{y}_d,\bm{\hat{y}}_d)
+\gamma \mathcal L_P(\bm{\theta})+\lambda \mathcal L_{Phy}(\bm{x},\bm{u},\hat{y})
\end{aligned}
\label{eq:phyinformL}\end{equation}
%\begin{equation}
%\mathcal{C}_{PhyIEQ}(\bm{z,u,\hat{y}}) \leq \zeta
%\label{eq:phyinformC1}\end{equation}
%\begin{equation}
%\mathcal{C}_{PhyEQ}(\bm{z,u,\hat{y}}) = \eta
%\label{eq:phyinformC2}\end{equation}\label{eq:phyinform}\end{subequations}
\end{small}\noindent
where $\mathcal L$ is the conventional loss function that measures the
distance between the predicted output and the target output. $\gamma$ and $\lambda$ are coefficients to weigh the regularization terms. $\mathcal L_P$ is the parametric regularization term (e.g., L1, L2, Tikhonov) to constrain the model complexity by regularizing the model parameters $\bm{\theta}$.
%$\bm{A},\bm{B}$ and $\bm{C}$. 
$\mathcal L_{Phy}$  is the physics-informed regularization term to add extra physics-driven constraints. Alternatively, the physics-informed information can be added as the inequality and equality constraints $\mathcal{C}_{PhyIEQ}$ and $\mathcal{C}_{PhyEQ}$ as follows:
\begin{small}
\begin{subequations}
\begin{equation}
\mathcal{L} = \mathcal L(\bm{y}_d,\bm{\hat{y}}_d)+\gamma \mathcal L_P(\bm{\theta} )\end{equation}
\begin{equation}
\mathcal{C}_{PhyIEQ}(\bm{x,u,\hat{y}}) \leq \bm{\zeta}
\end{equation}
\begin{equation}
\mathcal{C}_{PhyEQ}(\bm{x,u,\hat{y}}) = \bm{\eta}
\label{eq:phyinformC2}\end{equation}\label{eq:phyinform}\end{subequations}
\end{small}\noindent
where $\bm{\zeta}$ and $\bm{\eta}$ are arbitrary values to constrain the physics information. 

%\color{black}
The authors in \cite{Gupta2022} apply an adapted deep RL method with the training that is based on (\ref{eq:phyinform}) in wide-area damping control. The framework is called bounded exploratory control-based DDPG (deep deterministic policy gradient), consisting of NNs and polytopic controllers designed with the help of linear matrix inequality (LMI)-based mixed $H_2/H_\infty$ optimization. With the help of $H_2/H_\infty$-oriented stability physical constraints, the method is demonstrated effective 
%for both designs 
for both low-stressed and high-stressed networks.
%\color{red} check the last sentence. what is the physics-informed training in this work? \color{black}
%\color{black}
%\cite{Zheng2022} NN with the roubstfying terms
Universal learning machines such as NN can be informed of physical ingredients in training. The authors in \cite{Stiasny2021} use NN to realize a fast dynamic state and parameter estimation tool to assess power system stability. 
%The NN is trained using PMU data and the well-trained NN then can accurately describe the frequency dynamics of power systems. 
%The outputs of NNs are then connected to automatic differentiation (AD) where one can obtain the derivatives.
%The NNs are used to approximate physical differential and algebraic equations of power grids rather than learning the black-box mapping based on input-output values. 
The cost function for NN training is in the form of (\ref{eq:phyinformL}), consisting of a physics-informed regularization term based on the swing equation with respect to the NN output.  
%As such, \color{black} the NN tends to obey the underlying physical differential and algebraic equations of power grids by increasing the weights for the physics-informed regularization term. %The results show that the magnitude and type of noise have a minor impact, indicating the effectiveness of adding the physics-informed regularization.
%The authors in \cite{9282004} 
Similarly, the physics-informed NNs are used in \cite{misyris2021capturing} to capture the dynamics of power systems, with a group of state constraints and physics-informed differential and algebraic equation constraints incorporated in the NN architecture and loss functions equivalent to the form of (\ref{eq:phyinform}).
%The results show that the proposed physics-informed NNs can yield an accurate estimation of system dynamic response.
The authors in \cite{Li2021} apply NN for high impedance fault detection in power grids.  %Specifically, a convolutional auto-encoder is proposed to describe the voltage time-series during normal operations whereby one can detect high impedance faults. 
To obtain more reasonable NN parameters, \color{black}an elliptical regularization term is added in the form of (\ref{eq:phyinformL}) to incorporate the physics-regulated elliptical characteristics of voltages and currents into the cost function for training. %It shows that the training embedded with the hidden physics law in the regularization term enhances the detection performance even for noisy and non-full-observable scenarios. 
In summary, these physics-informed learning-based NNs show enhanced modeling efficacy and thus potentially beneficial for data-driven control on top of these NNs. %\color{red} maybe applications can be better specified when introducing these works

\color{black}

\subsubsection{Physics-Informed Model Structure Design}
%[Koopman][SINDY].
The physics-informed training discussed in Section B intends to incorporate physical system information in the cost function $\mathcal L$, while the physical information can also be integrated into model structure design, i.e., ``ML model'' in Fig. \ref{fig:SL}-\ref{fig:SL2}. In fact, proper modeling with the physics-informed structure design can nail down the learning space within reasonable areas, which could make the modeling more reliable and the learning faster.
%\color{red} check for the reduncant sentense \color{black}
%For example,
%\textbf{\emph{Pure Machine Learning Combined with Physics-Informed Structure Design}}.
Pure machine learning models as shown in Fig. \ref{fig:SL} and Fig. \ref{fig:SL2} can be incorporated with certain levels of physics interpretations in the structure. 
%For example, 
%"Bayesian RL provides an elegant approach to convey physical priors via value function or policy class in model-free RL ."
%to enhance the reliability and interpretability of RL,  the Markov decision processes can be constrained with Bayesian to include physical priors through value or policy functions \cite{ghavamzadeh2015bayesian,brunke2022safe}. %\color{red} It seems to be similar to the previous section? Or we may briefly explain at the beginning of this section that how it is different between physics-informed training and model structure. \color{black} 
%\color{black}
%The physics-informed NN %NN 
%structure design is a promising direction. 
For example, the authors in \cite{9282004} proposed a physics-informed NN %\color{red} as ``NN'' has been used to describe ``neural network'', please use NN after the definition throughout the paper (done) \color{black} 
tailored by automatic differentiation \cite{baydin2018automatic} of the original NN, whereby power system physical laws (e.g., underlying swing equations) are incorporated with the bounded space of admissible solutions to the NN parameters \cite{9282004}.
%\color{red} check these papers Nellikkath, R., & Chatzivasileiadis, S. (2022). Physics-informed NNs for ac optimal power flow. Electric Power Systems Research, 212, 108412. Zhao, Shuai, et al. "Parameter estimation of power electronic converters with physics-informed machine learning." IEEE Transactions on Power Electronics 37.10 (2022): 11567-11578. [Done. See \cite{nellikkath2022physics,zhao2022parameter} below,] \color{black}
 It has been shown that the physics-informed neural nets can effectively describe the rotor angle and frequency for uncertain power inputs and identify system inertia and damping parameters. Such design also provides opportunities to lower the requirement on the sizes of NNs and the training data set. 
 \color{black}
The physics-informed NN \cite{ostrometzky2019physics} that incorporates the regularization of power flow equations is proposed for power system state estimation under partial observability, achieving more accurate voltage-phasors estimation than conventional weighted least squares-based estimation.
The authors in \cite{wang2020physics} proposed a hybrid learning structure by incorporating the physical AC power flow model into the deep learning with autoencoders.  \color{black}
 The authors in \cite{nellikkath2022physics} use physics-informed NNs to accurately estimate the optimal power flow (OPF) with rigorous performance guarantee. The physical interpretation is further enhanced  by adding the disparities of AC-OPF Karush–Kuhn–Tucker (KKT) condition to the NN training loss function. 
\color{black} Additionally, the authors in \cite{zhao2022parameter} use physics-informed NN with an implicit Runge-Kutta integration \cite{stiasny2021learning} %NNs 
 at the local converter level, whereby deep NNs and converter physical models are seamlessly coupled and reliable parameter estimation of power converters are achieved. \color{black}

\subsection{Koopman-Based Methods}%Operator Theory}
%\textbf{\emph{Koopman-Based Methods}}. %Operator Theory}
%\color{red}
%Also, I feel that the advantage of Koopman-based methods---requires no offline training need to be mentioned somewhere. (done) 
%\color{black}

\color{black}
%The physics-informed rules and laws (physics-informed loss function, constraints, architecture design, etc.) become a growing consensus \cite{Huang2022RW} in terms of reducing data-driven modeling uncertainty and make the adopted learning machines more physically consistent,  generalized and interpretable. The fusion of reinforcement learning framework and physics-inspired information is another promising solution that needs attention for data-driven control, when considering the intrinsic connection between machine learning and control theories. The fusion can be conducted through catering the physics-informed rules and laws of supervised/unsupervised learning for reinforcement learning in environment surrogate model, value function or policy design \cite{Huang2022RW,Gros2020,ghavamzadeh2015bayesian,brunke2022safe}. Nonetheless, reinforcement learning based methods still suffers from some practical issues for power grid applications, such as the lack of scalability, adaptiveness and training efficiency \cite{She2022}. 

As aforementioned in Section 2, data availability, training efficiency, interpretability, adaptiveness, and scalability issues further motivate recent research works on adaptive Koopman operator control \cite{Gong2022,gong2023novel}, aiming to adaptively map nonlinear control to linear control that works for both small and large signals. 
%Specifically, the methods adopt small-signal linear space augmented with nonlinear bases and are adaptively identified to address time-varying uncertainty. Although these methods are online and can be applied without warm-up training, Koopman state space are determined empirically. Koopman generators that can estimate optimal and physical-consistent Koopman operators (based on physical states and control inputs) could be exploited with physics-informed machine learning and adequate offline data.
\color{black}
%Another emerging category of data-driven control is Koopman-based methods. \color{black} 
Koopman operator theory \cite{Proctor2018} %interprets nonlinear dynamics in a linear fashion, i.e., 
shows that a nonlinear dynamical system can be transformed into an infinite-dimensional linear system under a Koopman embedding mapping. The Koopman-enabled linear model structure is a way to interpret nonlinear dynamics, and is valid for global nonlinearity with the infinite-dimensional representation as opposed to traditional locally linearized small-signal models. Practically, one can consider finite-dimensional Koopman invariant subspaces where dominant dynamics can be described. In Koopman-based data-driven control, the physics interpretation on power grids of interest can be incorporated through the model structure design.
%different aspects\color{red} aspects?\color{black} including the structure, observable function selection, identification and control. 
%\textbf{\emph{Physics-informed Koopman-based structure}}. 
Particularly, given a nonlinear dynamical system with external control $\bm{x}_{k+1}=F(\bm{x}_{k},\bm{u}_{k})$, where $\bm{x} \in \mathcal{M}$ and $\bm{u} \in \mathcal{U}$ with $\mathcal{M}$ and $\mathcal{U}$ being the manifolds of state and control input, we consider the Koopman embedding mapping $\bm{\Phi}$ from the two manifolds to a new Hilbert space $\bm\Phi: \mathcal{M} \times \mathcal{U} \to \mathcal{H}$, which lies within the span of the eigenfunctions $\varphi_{j}$. That is, $\bm{\Phi(x,u)}= \sum_{j=1}^{N_{\varphi}}\varphi_{j}(\bm{\bm{x},\bm{u}})\bm{v}_{j}$, where $\bm{\Phi}(\bm{x},\bm{u})=[\Phi_{1} (\bm{x,u}),\Phi_2(\bm{x,u}),…,\Phi_i(\bm{x,u}),…,\Phi_p(\bm{x,u})]^T$ is a set of Koopman observables, $\bm{v}_{j}$ are the vector-valued coefficients called Koopman modes. %and $N_{\psi}$ is the number of eigenfunctions. 
The Koopman operator $\mathcal{K}$, acting on the span of $\varphi_{j}$, advances the embeddings $\bm{\Phi(x,u)}$ linearly in the Hilbert space $\mathcal{H}$ as \cite{Proctor2018}:
\begin{small}
\begin{equation}
\begin{aligned}
\bm{\Phi}(\bm{x}_{k+1},\bm{u}_{k+1}) & =\mathcal{K}\Phi(\bm{x}_k,\bm{u}_k)\\
& = \mathcal{K}\sum_{j=1}^{N_\varphi}\varphi_j({\bm{x},\bm{u}})\bm{v}_j
= \sum_{j=1}^{N_\varphi}(\rho_{j}\varphi_j(\bm{x}_k,\bm{u}_k)\bm{v}_j)
 \end{aligned} \label{eq:koc}
\end{equation}\end{small}\noindent
%\color{red} can be written in a more concise way \color{black}
where $\rho_{j}$ are the eigenvalues satisfying $\mathcal{K} \varphi_j(\bm{x,u})=\rho_j \varphi_j(\bm{x,u})$. 
%\subsubsection{Extended Dynamic Mode Decomposition with Control}
In (\ref{eq:koc}), one can assume that $\Phi_i(\bm{x,u})=g_{i}(\bm{x})+l_i(\bm{u})$ where $g_{i}(\bm{x})$ is a nonlinear observable function and  $l_i(\bm{u})$ is linear with $l_i(\bm{0})=0$ %\color{red} one is uppercase and one is lowercase? \color{black} 
\cite{KORDALinear2018}. In addition, we assume $\Phi_{i}(\bm{x}_{k+1},\bm{0}) = \mathcal{K} \Phi_{i}(\bm{x}_k,\bm{u}_k)$ for all $k$. Then, $g_{i}(\bm{x}_{k+1})+l_i(\bm{0})=\mathcal{K} g_{i}(\bm{x}_k)+\mathcal{K} l_{i}(\bm{u}_k) \Rightarrow g_{i}(\bm{x}_{k+1})=\mathcal{K} g_{i}(\bm{x}_k)+\mathcal{K} l_i(\bm{u}_k)$. This assumption means that the Koopman operator is only attempting to propagate the observable functions at the current state $\bm{x}_k$ and inputs $\bm{u}_k$ to the future observable functions on the state $\bm{x}_{k+1}$ but not on future inputs $\bm{u}_{k+1}$ \cite{Proctor2018}. 
%(i.e., $[\bm{\Delta P}^*,\bm{\Delta Q}^*} ]^T$ are not state-dependent
%\color{red} we didn't mention $[\bm{\Delta P}^*,\bm{\Delta Q}^{*} ]^T$ \color{black} )\color{red} Are these descriptions about Koopman operator theory generic enough? I suppose it is only for our works. Do they also apply to other works? e.g., Power grid transient stabilization using Koopman model predictive control\cite{KORDA2018297}\color{black} (it works for \cite{KORDA2018297} as this paper also assumes the same EDMD form, which is a simplified approximation of Koopman operator. The general Koopman operator theory however does not inlcude the assumption, but it is required to simplify Koopman operators to EDMD) \color{black} . 
Let us define $\bm{z}:=\bm{g}(\bm{x})=[g_1(\bm{x}),g_2(\bm{x}),…,g_i(\bm{x}),…g_p(\bm{x})]^T$. Then we have an approximation of (\ref{eq:koc}) in a form of extended dynamic mode decomposition with control (EDMDc) \cite{KORDALinear2018} as below %\color{black}
%\begin{small}
%\begin{subequations}
%\begin{equation}
%(Process \quad  model) \quad \bm{z}_{k+1} = %\bm{A}\bm{z}_k+\bm{B}\bm{u}_k+\bm{\delta}_k
%\label{eq:process}\end{equation}
%\begin{equation}
%(Observation \quad  model) \qquad  %\bm{y}_k := \bm{x}_k = %\bm{C}\bm{z}_k+\bm{e}_k
%\label{eq:observation}
%\end{equation}\end{subequations}\end{small}\noindent
%\cite{KORDALinear2018} as below
\begin{small}
\begin{subequations}
\begin{equation}
(Process \quad  model) \quad \bm{z}_{k+1} = \bm{A}\bm{z}_k+\bm{B}\bm{u}_k+\bm{\delta}_k
\label{eq:eDMD1}\end{equation}
\begin{equation}
(Observation \quad  model) \qquad  \bm{y}_k = \bm{C}\bm{z}_k+\bm{e}_k
\label{eq:eDMD2}
\end{equation}\end{subequations}\end{small}\noindent
where $\bm{y}_k$ are the outputs of the Koopman state space model. 
%\color{black}
%\textbf{\emph{Sparse Identification of nonlinear dynamics with control (SINDYc)}}.
The Koopman observables in (\ref{eq:eDMD1})-(\ref{eq:eDMD2}) can integrate power system physical domain knowledge %be conducted empirically based on domain knowledge 
\cite{brunton2022data,KORDA2018297,Netto2021,Gong2022,husham2022decentralized}. For example, the authors in \cite{KORDA2018297} selected the Koopman observables with osillotary terms ($\sin\theta$ and $\cos\theta$) to address sinusoidal-driven interaction dynamics that emerge when subject to large perturbations and low inertia. These trigonometric terms are physics-informed ingredients because the general solution for differential power flow equations contains trigonometric patterns. Likewise, the authors in \cite{Gong2022,Netto2021} also include the functions $\sin\theta$ and $\cos\theta$ into the Koopman embedding to describe underlying power flow interaction dynamics.
%\color{red} I think 2 -3 sentences may be added to further explain how the physic information is integrated in these works as this section intends to discuss physics-informed design. (done).
%\color{black} 
Besides the interpretability of dynamics, another advantage of Koopman-based methods is the linearity enabled by Koopman operators and therefore requires no offline learning, enabling online adaptiveness. \color{black}
Learning-based methods using deep auto-encoder, deep NNs, automated dictionary learning \cite{lusch_deep_2018,Morton2019,Han2020,chen_variants_2012,Yeung2019} are also employed to help determine Koopman state spaces, whereas the optimal discovery of Koopman embedding mapping remains an open question due to complex nonlinear systems of high dimensions and uncertainty.
\color{black}
Given the Koopman model state space, the model parameters can then be estimated online by least-squares regression \cite{KORDALinear2018, KORDA2018297,husham2022decentralized}, iterative learning \cite{Gong2022}, enhanced OKID \cite{gong2023novel}, etc.\color{black}
%\color{red} how does it (finding Koopman embedding mapping challenging) relate to SINDy? The logic is not very clear. 

Sparse identification of nonlinear dynamics (SINDy) is another type of method for the data-driven discovery of dynamical systems characterized by sparse dominant features. We classify SINDy as a type of Koopman-based method as it has been proved in \cite{klus2020data} that SINDy is a special case of EDMD. As dynamical systems can be often described by a few governing equations, SINDy can be extended to discover leading Koopman eigenfunctions whereby realizing the control on the discovered spans \cite{kaiser2018discovering, kaiser2021}.
\begin{small}
%\begin{subequations}
\begin{equation}
\bm{\dot \bm X} = \bm{\Xi}\bm{\Theta}^T(\bm{X},\bm{U})
\label{eq:sindy}\end{equation}
%\begin{equation}
%\bm{y}_k = \bm{C}\bm{z}_k+\bm{e}_k
%\label{eq:Koopmanobservation}
%\end{equation}
%\end{subequations}
\end{small}\noindent
where $\bm X$ and $\bm U$ are the state data matrix and control input data matrix, respectively. $\bm{\Xi}$ is the coefficient matrix to be identified with sparsity. $\bm{\Theta}$ is a library of candidate nonlinear functions that compose the potential feature space to be identified. The model identified with SINDy is similar to (\ref{eq:eDMD1}) in the sense that both describe the nonlinear dynamical evolving in a linear fashion on the span of \color{black}a few nonlinear features, %\color{red} this is not very clear \color{black}, 
which enables the use of mature linear control with well-characterized stability. The proper selection of the feature candidates can inform the model of physics. For example, the authors in \cite{cai2022} use SINDy to locate forced oscillation sources. Based on the physical formulation of stochastic power system dynamic model \cite{cai2022}, the power system dynamics can be seen sparse in the feature space of zero-degree polynomial bias, linear functions and trigonometric functions.
% \color{red} similarly, we may explain briefly how the physical information is considered by SINDy. (done)
%\color{black}

The Koopman-based methods have been investigated in many previous works in power systems, such as power system nonlinearity modeling, stability assessment, and forced oscillation location \cite{Netto2021, Yeung2019, cai2022}. 
Although the external control inputs are yet to be incorporated in these applications, it is natural that the power system control can be conducted based on the Koopman state space with mature linear control algorithms (e.g., optimal control, predictive control, or robust control). That being said, the nonlinear power system control can be converted to linear control in the lifted Koopman state space. For example, the LQR and model predictive control techniques have been applied on the Koopman model (\ref{eq:eDMD1})-(\ref{eq:eDMD2}) to realize microgrid voltage and frequency secondary control, transient frequency control \cite{Gong2022,KORDA2018297}. %etc. 
%Considering time-varying environment of power system operation, rolling-based identification for a predictor based on past information + optimization for now and future based on the predictor is favorable to further handle the time variance and nonlinearity.
%For example, The LQR and model predictive control (MPC) techniques have been applied on the Koopman model (\ref{eq:process})-(\ref{eq:observation}) to realize microgrid voltage and frequency secondary control, transient frequency control \cite{Gong2022,KORDA2018297}, etc. 
The authors in \cite{Ping2021} proposed %\color{red} the tense may need to be consistent. Past tense or present tense. I think past tense might be better but that means modifications in previous sections when introducing previous works are needed (yeah, details to be discussed) \color{black} 
deep Koopman model predictive control for improving transient stability and frequency regulation. Particularly, a deep NN was used to obtain more effective observable functions for Koopman state space with the training according to a Koopman-oriented cost function.
Likewise,  deep NNs were used in \cite{you2018deep} to learn a Koopman operator model in the presence of price spoofing in the context of market-based frequency regulation. %\color{red} Is physic information considered in these works? (No. TBD)\color{black} 
The learned Koopman model was used to realize a robust data-driven control against price proofing with Lyapunov-based LMI constraints to guarantee asymptotic stability, which enhanced cyber security.

\color{black} In summary, emerging Koopman-based methods are promising due to: %considering the following facts: 
(i) %the Koopman-based methods 
Dealing with nonlinear problems in a linear and dynamics-interpretable fashion, %which tends to yield 
leading to faster and optimal system identification with tractable identification process and well-characterized stability \cite{Gong2022}. 
%(ii) Because of the linearity and interpretability, the deployment can be more flexible for integrating physical information and facilitate scalability. 
(ii) %the Koopman-based methods can have 
Generating fixed states (%some people call them 
i.e., Koopman observables) from measured states that have explicit physical meaning, allowing for easy extension of conventional model predictive control frameworks to the Koopman state space \cite{KORDA2018297, Netto2021}. %which can be generated from and combined with the measured states that have explicit physical meaning, whereby conventional model predictive control framework can be easily extended to the Koopman state space \cite{KORDA2018297, Netto2021}. 
(iii) Separating the determination of the Koopman state space from parameter estimation, enabling online identification without warm-up training and facilitating fast adaptive control. %Koopman-based methods decompose the determination of the inherent static problem structure (i.e., Koopman state space) and the parameter estimation to separate tasks in an interpretable manner. %\color{red} not so sure if (iii) is needed as interpretability is mentioned in (i)? (this is to justify Koopman is promising in terms of online identification while (i) is to emphasize the physical consistency of model structure.\color{black} 
%Therefore, so long as the Koopman state space is determined, the Koopman-based model can be identified online without warm-up training and then fast adaptive control can be realized. 
%\color{red} come back again to see if these were discussed in future trends. (there might be a little repetition compared to Section 5.3: online identification for different categories of data-driven control. However, it looks fine as this section is method-oriented and details why the Koopman methods are promising from different perspectives. Section 5.3 discusses the online adaptiveness of Koopman-based method based on the fact (iv). Section 5.4.3 is about potential countermeasures based on the facts)\color{black} 

%\cite{DBLP:journals/corr/abs-1902-09742,Han2020,lusch_deep_2018,KORDA2018297,Netto2021,chen_variants_2012}]

%[An Adaptive Generalized Predictive Control Method for Nonlinear Systems Based on ANFIS and Multiple Models \cite{Zhang2010}], [Ensemble Koopman-Based Modeling\cite{Gong2022,Pastor2020}].

\hspace{-3cm}
\begin{figure}[!tb]
\centering \footnotesize \color{black}Table 2

\footnotesize \color{black} Summary on Data-Driven Control Methods and Applications in Power Grids

\centering
\includegraphics[width=1.0\linewidth]{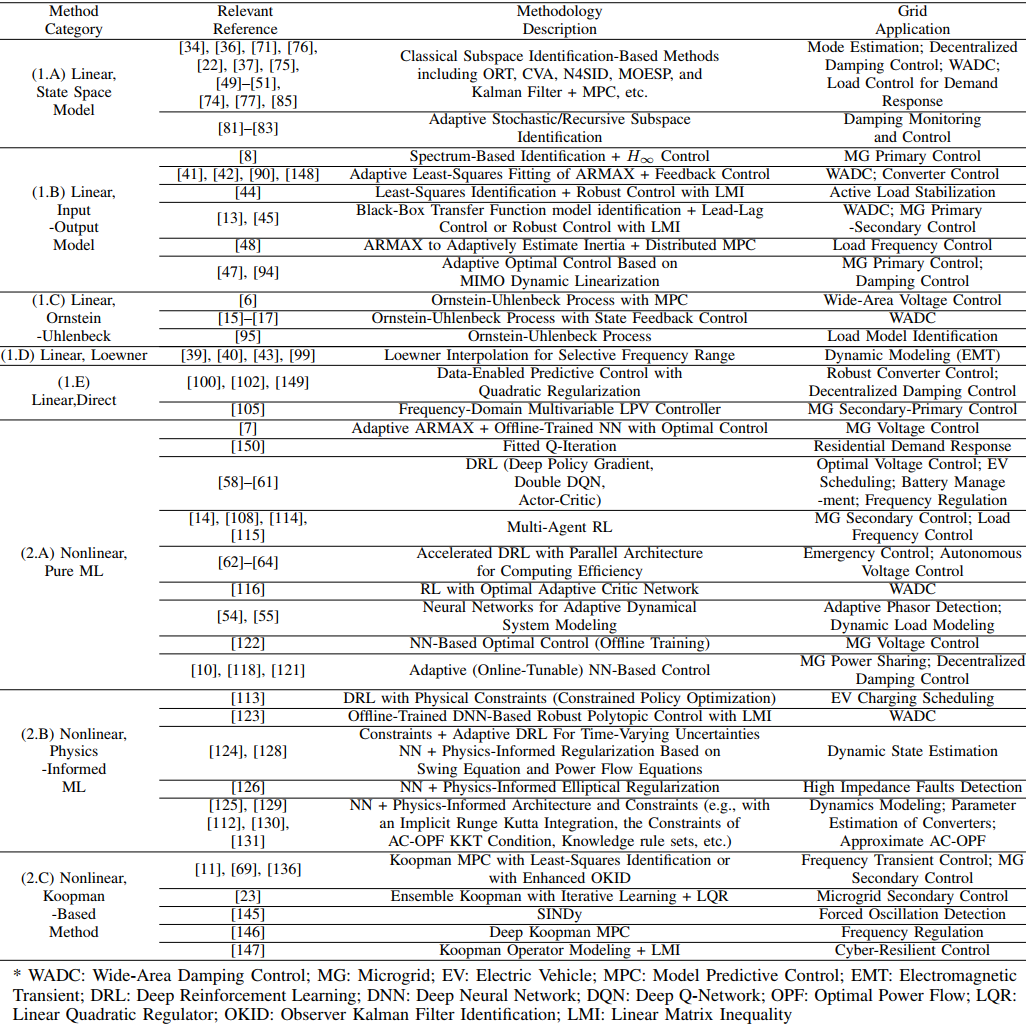}
  \label{tab:ddcontrol_PS}
\end{figure}

\section{Overview and Future Trends}
\color{black}
%Overall, we summarize the typical data-driven control methods discussed in Sections 3-4 and their applications in power grids in Table 2. Figure \ref{fig:method_IDandCTRL} also presents the methodology categories with the classification of data-driven identification and control.
%The methodology pros and cons, power system measurement data and operation requirements make the methods suitable for different applications.
%\color{red} this paragraph especially the last sentence doesn't read well. ChatGPT gives ``
In summary, the data-driven control methods discussed in Sections 3-4 and their applications in power grids are presented in Table 2. Figure 5 provides an overview of the methodology categories for data-driven identification and control. These methods offer various advantages and disadvantages, and are inherently suitable for different grid applications.
%their suitability for different applications is influenced by factors such as power system measurement data and operation requirements. 
%'' for your reference\color{black}

\begin{figure}
\centering
  \includegraphics[width=1.0\linewidth]{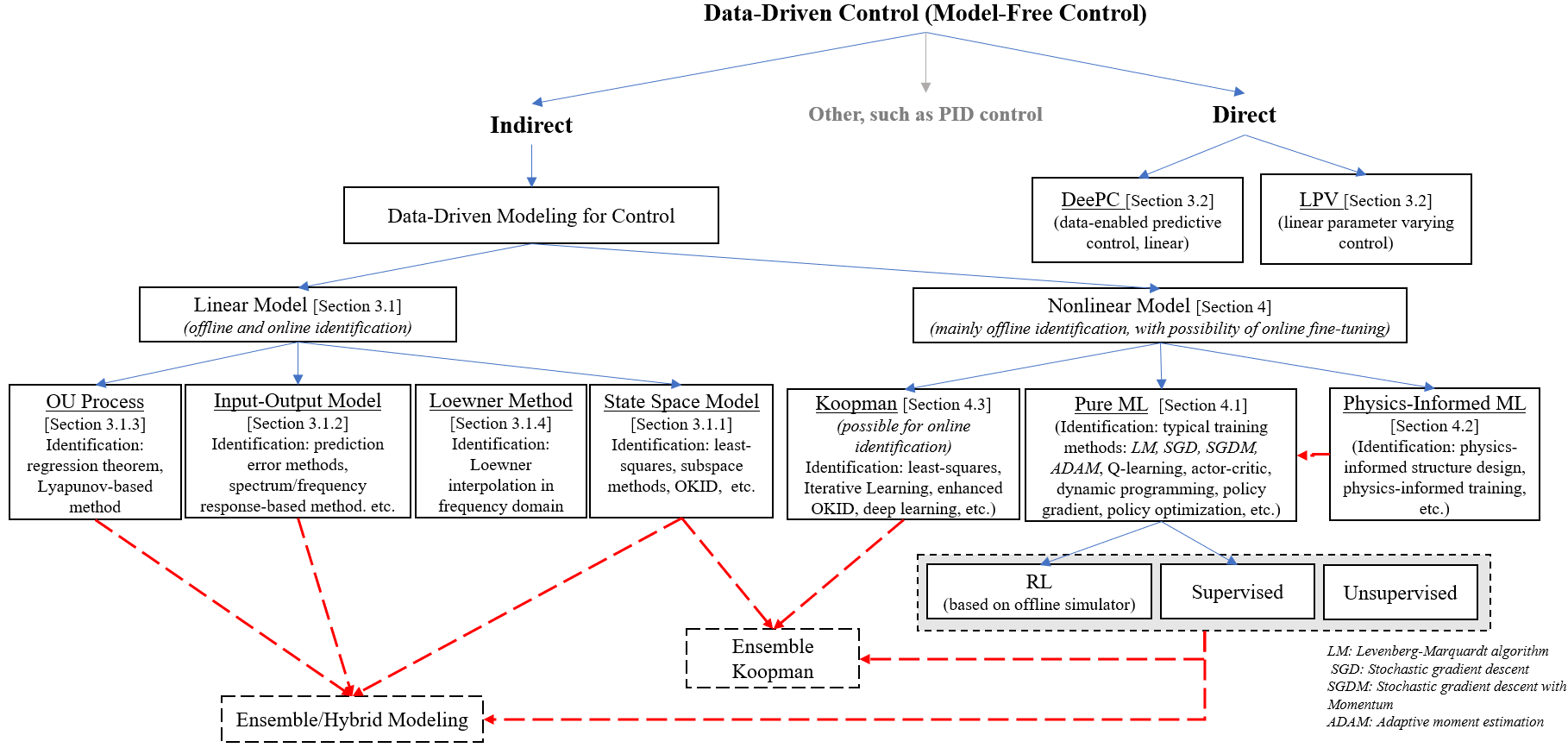}
  \caption{Methodology category for data-driven identification and control.}
  \label{fig:method_IDandCTRL}
\end{figure}

\subsection{Overview of Existing and Potential Power Grid Applications}
The data-driven methods are generic and can be tailored for various grid applications. According to the existing work listed in Table 2, we will discuss selective application scenarios of data-driven control. They are oscillation damping control, microgrid control, and aggregated load/EV control.

%[data availability, quantity, granularity and quality]

\color{red} 
%[The original thoughts to the question in red: less chance that the linear model is overfitting. 
%The key points include: (1) Linear model is suitable for damping control under small-signal assumption. (2) Online identification is feasible because of damping control requirements and the linear identification efficiency. (3)  Although there may be larger model uncertainty for converter control, a linear model can also be used in normal operation scenarios with adaptive linear modeling and robust control design.]

%[It seems load frequency control is also a popular application, as there are linear methods and also learning-based methods applied in the previous works. Also, are load frequency control and frequency regulation similar?]

%[some background information, then what the challenges are (e.g., uncertainty from end-user), the data time scale, and then what kind of methods have been proposed to address the challenges. ] 
\color{black} 

%\textbf{\emph{Decentralized damping control/Inverter control:}} 
\textbf{Oscillation damping control:}
%\color{red} must be decentralized? () 
\color{black} 
The first representative application in power grids is the damping control of low-frequency 
%\color{red}[electromechanical, i.e., low-frequency oscillation as well? (my impression is that subsynchronous oscillation doesn't include electromechanical oscillations)] \color{black} 
oscillations in either decentralized or centralized manners \cite{ramakrishna2010adaptive,Ilias2020}. As the small signal assumption holds in this application scenario, linear models are generally expressive enough to describe the oscillation modes.
%The increasing penetration of power converter-interfaced devices is different from traditional synchronous generators and thus changes the power grid dynamic characteristics. 
Diverse local devices, global network topology and parameters, as well as time-varying operating conditions nowadays make power grid oscillatory dynamics suffer from higher levels of uncertainty, compromising the performance of conventional model-based controllers. Fortunately, such uncertainty can be compensated by data-driven controllers in either a decentralized or centralized fashion. Examples of decentralized adaptive linear data-driven methods are with linear input-output transfer function model \cite{Ryan2021}), ARMA \cite{ramakrishna2010adaptive,shakeel2020line} and DeePC \cite{huang2019data, Huang2022}.
%Because of the diverse dynamic characteristics of local devices as well as the line topology and parameters, the systemwide stability of power grids become complex and local damping sometimes is not enough to suppress the oscillatory interactions. 
With the development of wide-area measurement infrastructure, centralized wide-area damping control can be applied to enhance the modeling and thus damping of systemwide interactive oscillation. 
For example, state space model with subspace identification \cite{Nieto2015,WideAreaControl_N4SID, YE2013509}, ARMAX \cite{Liu2017}, as well as Ornsein-Uhlenbeck process \cite{Georgia2021,Ilias2020,Guo2021} with online parameter estimation are used for damping control with wide-area phasor measurement units. Machine learning such as RL is also used in wide-area damping control to adaptively learn how to address nonlinearity and uncertainty \cite{Yousefian2018}. However, the performance heavily depends on its offline simulator and the adaptation is slow. Considering the simulator fidelity, offline training efficiency and optimality issues, the application purely with RL is not as practical as adaptive linear methods.

\textbf{Microgrid control:}
%\color{red} make sure if you focus on secondary control or not? (done) \color{black}
Microgrid control plays a vital role in %takes care of 
frequency and voltage regulation/restoration, as well as systemwide power and energy management. %The dynamic characteristics of microgrids vary depending on the specific distributed resources and networks involved. %The dynamic characteristics of microgrids depend on case-by-case distributed resources and networks. 
Roughly speaking, the dynamics in microgrids span from hundreds of milliseconds to minutes, and the data-driven controller can sample measurements in the range of tens of milliseconds to seconds. % the timescale of the dynamics ranges from hundreds of milliseconds to minutes, and the measurements could be sampled by the data-driven controller in the order of tens of milliseconds to seconds. 
Different from traditional power grids, microgrids are characterized by low inertia, coupled states, and wide-frequency dynamics response and are susceptible to high uncertainty and nonlinearity when confronting large disturbances.
To address these challenges, nonlinear learning-based methods, such as RL \cite{Younesi2020,KOU2020114772,Khooban2021,Chen2022}, NNs \cite{Ma2021,Zheng2022,dong2018artificial} and Koopman-based methods \cite{Gong2022,gong2023novel} have been employed.

\textbf{Load/EV control for demand response:} Another suitable application scenario of data-driven control in power grids are load and EV control for demand response.  The root causes of complicated power consumption dynamics are the combinations of device physical dynamics (e.g., thermal dynamics for thermostatic loads), end-user behavioral dynamics (population dynamics of aggregated load and EVs) as well as the electricity market dynamics.  %However, accessing or accurately sensing end-user information introduces uncertainty. 
Usually, the end-user information may not be accessed or accurately sensed, introducing uncertainty. 
The timescale of the aggregation dynamics typically ranges from minutes to hours or even days. The measurements could be sampled by the data-driven controller at the order of minutes from smart meters. From the technical control perspective, %\color{red} what do you mean by direct control ? (direct load control means the load can be controlled directly by the controller rather than indirectly controlled by prices or incentives. I deleted the word "direct" here to avoid terminology confusion with direct data-driven control) \color{black}
%a reduced-order state space model is well-suited for a large population of load or EVs to realize full responsiveness to control requests in a linear fashion \cite{Mathieu2012}
a reduced-order state space model is well-suited to efficiently describe the dynamic state evolving of a large population of load or EVs in a linear fashion. 
%\color{red}
This allows for the predictive controller design to realize] full responsiveness to control requests from system operators, while considering the constraints imposed by the normal end use of individual loads/EVs \cite{Mathieu2012}.
\color{black}
%whereby the full responsiveness to the control requests of system operators can be realized by linear control design considering the constraints of normal end use of individual loads/EVs \cite{Mathieu2012}. %\color{red} what does it mean by "full responsiveness to control requests in a linear fashion"? (it has been rewritten)\color{black}. 
%\color{red}
%(Is this the same as what you want to say? Pretty similar while I add `the predictive controller design to realize') 
%\color{black}
Online identification on top of the reduced order model is favorable to adapt to the changes associated with weather, financial or social factors that may affect end users. From the commercial operation perspective (e.g., economic dispatch, power tracking quality, grid service provision), nonlinear methods such as RL \cite{Bui2020,Li2020, Liu2019, Ruelens2017} can directly address the nonlinearity of electricity market dynamics and end-user behaviors. %the nonlinearity of electricity market dynamics and end-user behaviors can be directly addressed in nonlinear methods such as reinforcement learning \cite{Bui2020,Li2020, Liu2019, Ruelens2017}. 
Usually, stable operation over the timescale of hours is the prerequisite %before designing data-driven methods 
to investigate optimal scheduling and operation. %The system can operate normally even without control. 
Therefore, the application of machine learning is practical in this context, as training efficiency is usually not the primary concern. %Thus, the application of machine learning is practical as training efficiency is usually not the main obstacle.

%\color{red} [It seems load frequency control is also a popular application, as there are linear methods and also learning-based methods applied in the previous works. Also, are load frequency control and frequency regulation similar?][Also, I suppose you want to explain dynamics time scale and data for all applications, otherwise, it doesn't seem necessary unless it was asked by reviewers]
%\color{black} \textbf{\emph{Load Frequency Control:}} (TBD if we need to add this or only keep three selective applications)

\subsection{Practicality Overview of Data-Driven Control Methods}
The data-driven identification and control can be conducted based on three types of measurement data (ambient, transient, and probing data) \cite{Liu2017}. Among them, the ambient data and
transient data are suitable for online applications when subject to ambient conditions and large disturbances, respectively. The probing data means actively injecting a probing signal into the system. The signal needs to consistently excite the system to obtain informative datasets, and the probing can deteriorate the electricity quality of power grids.
The data availability, granularity and quantity highly depend on the grid application scenarios and the methods adopted. 
\color{black} Use the three selective grid applications discussed in Section 5.1 as examples. They usually adopt different identification and control methods and thus have different data requirements. Linear models are well-suited for %decentralized 
oscillation damping control, %\color{red} need modification given the modified application? (done) \color{black}, 
while nonlinear methods are gaining popularity for microgrid control and demand response. Linear models typically require a relatively small volume of data. For example, in wide-area damping control, accurate identification of linear models can be achieved with online data consisting of a window length of 180-300 seconds of ambient data \cite{Ilias2020,Liu2017} or 10-second ring-down data \cite{Liu2017}  sampled at a frequency of 30Hz from wide-area synchrophasor measurements.
On the other hand, %with the increasing  penetration of diverse power converter-based resources in power grids, the interaction dynamics become more complex and data-driven nonlinear modeling and control gain attention. 
the nonlinear methods usually need to train a model offline with large volumes of data (e.g., tens of thousands of examples or more) to obtain a deterministic warm-up reference, which then can be fine-tuned online with a small piece of data to adapt to varying uncertainty. 
\color{black} Koopman-based methods have the potential to use a small volume of data that are comparable to linear models without the need of warm-up training, thus having the potential to realize faster and more accurate adaptation for grid nonlinear dynamics provided that the Koopman state space (approximating the Koopman embedding mapping) is properly predetermined.

The quality requirements of data are mainly determined by the preprocessing techniques and the data-driven identification and control methods employed. %The data quantity and quality requirements depend on data preprocessing as well as data-driven identification and control methods. 
Even within the same method category, different algorithms for preprocessing, identification, and control may have varying data requirements based on the application scenario. Thus, it is challenging to establish exact requirements that are universally applicable. %different algorithms of data preprocessing, identification and control may correspond to different data requirements in different application scenarios. Therefore, it is difficult to have exact requirements that apply to all.
Data preprocessing can involve proper filtering and interpolation to improve data quality. 
%Proper filtering and interpolation can be conducted in data preprocessing to help improve the data quality. 
Stochastic optimal identification (e.g., OU process regression theorem, Kalman filtering) and robust control design (e.g., $H_\infty$, LMI) can also be incorporated to reduce the requirements on data quality. Table 3 provides a brief overview of commonly used models and their corresponding data requirements for three selected grid application scenarios. 

\begin{figure}[!tb]
\centering \footnotesize \color{black}Table 3

\footnotesize Summary on Dominant Modeling Methods and Data Requirement 

for Selective Grid Applications

\centering
\includegraphics[width=1.0\linewidth]{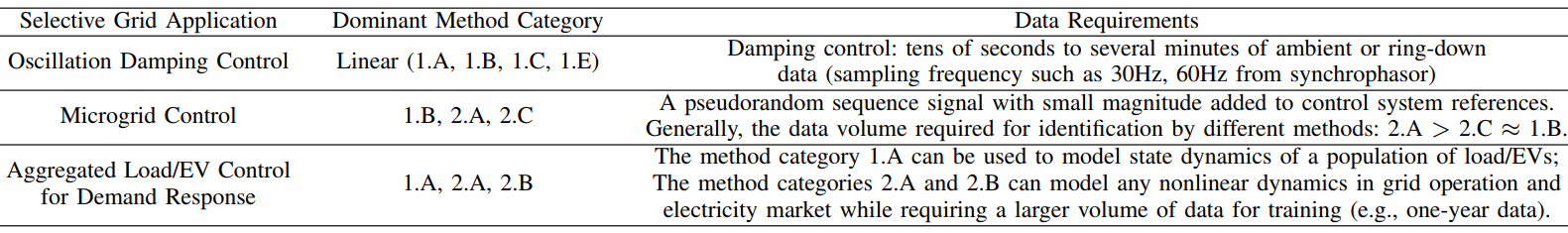}
  %\label{tab:GridApp_DataReq}
\end{figure}

\color{black}
\subsection{Online Identification for Different Categories of Data-Driven Control}
%\color{red} First, I think it may not be discussed in the section of ``Future trends". It may be moved to the place after 5.2. \color{black}
Online adaptiveness is a desirable feature in power systems due to the time-varying nonlinearity and complexity they exhibit. %Obtaining an offline model that consistently performs well is challenging in such dynamic environments. 
By utilizing a small amount of online data, online identification effectively addresses uncertainty and mitigates modeling issues. 
Online identification can be integrated into all categories of data-driven control methods, though it may be inherently easier in some categories compared to others. 
%Online adaptiveness is a favorably trending feature in power systems. This is so because the time-varying nonlinearity and high complexity of power grids pose challenges to obtaining an offline model that is always effective. Adaptive identification based on a relatively small piece of online data could be beneficial to quickly bound uncertainty propagation, and thus mitigates the modeling problem. Online identification can be incorporated into each category of data-driven control methods, while for some categories it may be inherently easier than others.

\textbf{For linear data-driven identification and control}. 
%\color{red} (modified)
\color{black} Generally, most of the linear data-driven identification and control methods discussed in Section 3 are more computationally efficient than nonlinear methods, thus they can realize online identification and control by using rolling windows, so long as the identification at each time step can be completed within a time interval specified for the grid application of interest. For example, the rolling window-based online identification has been used to online identify OU process models for wide-area voltage control \cite{Georgia2021} and dynamic load modeling \cite{Pierrou2020}. In some cases, recursive formulations can further improve the algorithms' efficiency. For example, the recursive or stochastic subspace methods have been used to online identify state space model for adaptive damping control \cite{wu2006multivariable,zhang2012adaptive} and electromechanical mode estimation \cite{sarmadi2013electromechanicalf}. The recursive least-squares method has been used to online identify linear ARMAX models for adaptive control of a converter in a grid-tied microgrid \cite{shakeel2020line} and of power system stabilizer \cite{ramakrishna2010adaptive}. The recursive estimations of some matrix quantities \cite{Sheng2020} have been used to online identify OU process linear models for the estimation of dynamic system state matrix of multi-machine power grids.
%(please check and modify as you see appropriate) 
\color{black} %Adaptiveness is relatively easy to obtain as the linear methods tend to require fewer data to conduct the identification and have higher identification efficiency compared to nonlinear methods. Adaptive control based on the black-box linear input-output model ARMA and the state space representation has been successfully applied to adaptive inverter control \cite{shakeel2020line} and power system stabilizer control \cite{wu2006multivariable}. Although power systems are high-order nonlinear systems, the adaptive linear identification and control are well suited when linearized approximation, to some extent, can faithfully represent the essential features of the system.

\textbf{For machine learning-based nonlinear identification}. The idea of pre-training and online tuning was explored for power system analysis in nineties \cite{Kamwa1996RNN}. In 1999, adaptive neuro-identifier and -controller were proposed in \cite{shamsollahi1999application} for power system stabilizer in a multi-machine power system. These approaches involved obtaining a pre-trained machine learning model through offline warm-up training, which served as a deterministic baseline for subsequent online tuning. This baseline helped reduce output oscillation \cite{shamsollahi1999application}. %Similarly, the pre-trained machine learning model was obtained with offline warm-up training, and then was used as a deterministic baseline on top of which the online tuning was conducted. (An accurate baseline can help reduce output oscillation \cite{shamsollahi1999application}.) 
Nonetheless, the reliability of adaptive machine learning-based methods is limited due to the lack of physical interpretability and the biased information from online data. %for universal learning machines. %the adaptive machine learning-based methods are not reliable due to the lack of physical interpretability, as well as the biased information of online data for universal learning machines.
Deserved to be mentioned, such an offline-training online-tuning paradigm  aligns with the concept of  transferable features and learning in the machine learning community, which aims to transfer the knowledge learned from existing system data to new similar systems through offline training, with fine-tuning of learning machine parameters %are fine-tuned 
when new data is available \cite{yosinski2014transferable}.
The development of transfer learning techniques nowadays \cite{vrbanvcivc2020transfer} may provide additional opportunities for researchers in power sectors to borrow up-to-date ideas from machine learning community to realize more time-efficient and reliable adaptive learning for nonlinear data-driven control, in terms of trade-offing present and past data in a rolling and more intelligent fashion. %\color{red} is transfer learning a new trend of supervised learning? also for Section I. we will discuss it.(modified) \color{black}

\textbf{For Koopman-based methods.} %\color{red} I will check this part later \color{black} 
To our best knowledge, Koopman-based models provide an opportunity to realize online nonlinear identification.
%while in the early stages. 
They are well suited for online identification by nature due to the linearity after Koopman embedding mapping. They are particularly favorable when obtaining a ``warm-up" model is challenging due to limited on-field data volume and runtime requirements. %The Koopman-based methods are also favorable when obtaining a ``warm-up" model to start with is problematic due to the lack of a large volume of on-field data and the run time requirements. 
For example, Koopman-based microgrid secondary voltage and frequency control \cite{Gong2022,gong2023novel} were successfully implemented without warm-up training, even when the microgrid configuration and control parameters were unknown. %conducted without warm-up training in the scenario that the microgrid configuration and control parameters are unknown. 
However, the determination of the Koopman state space still relies on empirical methods. An alternative approach is to pre-learn Koopman generators using physics-informed machine learning and offline data in power grid control. This allows for estimating optimal and physically consistent Koopman operators based on physical states and control inputs. Subsequently, the parameters of the Koopman state space model can be identified online in a linear identification manner.
\color{black}

\subsection{Future Trend}
Modern power systems have increasingly high uncertainty and nonlinearity due to the increasing penetration of inverter-based resources, leading to more complex system dynamics within wider frequency bands than conventional power grids. As a result,  modeling efficacy %\color{red} or accuracy? 
%\color{black} 
tends to be more problematic. %The modern power grid dynamics such as electro-mechanical and electromagnetic transient (EMT) are at different time scales with overlaps from high-order harmonics to super/sub-synchronous dynamics. 
%The control at different time scales is coupled and needs to be seamlessly designed against nonlinearity and uncertainty in environments. Fast dynamics need to be handled quickly by robust control; the others have looser requirements on the response time %speed 
%but need to ensure long-term safety and operation optimality. %to handle the time scale overlaps of control systems. 
%For example, the authors in \cite{cao2021deep} propose RL-based two-timescale voltage control for active distribution systems. For the fast time scale, the power grid is partitioned into several subsystems based on voltage-reactive power sensitivity; then the scheduling of PV inverters in the whole grid is treated as multi-agent Markov games with each subsystem modeled as an intelligent agent. For the slower time scale, a single soft actor-critic algorithm with global information is used to coordinate on-load tap changer and switched capacitors while considering control behaviors of the inverters. 
Conventional machine learning-based methods may not be able to realize satisfactory control performance efficiently and safely due to increasingly high complexity of power grids and the ``black-box" nature of machine learning. Emerging data-driven control methods %is required 
could better address uncertainty and nonlinearity, while being interpretable or having certain physics-informed performance guarantees to enable their applications in real-world power systems. 
%\color{black} Some of the emerging data-driven control methods can adapt to time-varying operating conditions via online identification/learning, which becomes favorable with the evolving of grid modernization. 
%\color{red} why is the prevoius sentence needed? (deleted. Previously there was a section about online identification in future trend. Now it has been moved to Section 5.3)
\color{black}  In what follows, we will briefly discuss the trend of each method category. 
%\color{black} Specifically, data-driven control with machine learning-based are suitable for slow operational dynamics. Although the training of advanced learning machines are time-consuming, the powerful nonlinear fitting and learning-based optimization capability that can reduce operational uncertainty. Data-driven control with Koopman-based methods could be further applied for control at fast time scales due to its inherent interpretability on dynamics and the fast online adaptiveness enabled by Koopman-based linear identification.
%\color{red} will the following discussions address the two-time scale issues? (done) \color{black}%TBD. Shall we give a brief summary about the issues to be discussed next? \color{black} %applications safely and reliably.
%However, there are a few practical challenges that need to be addressed such as the challenges regarding data collection, safety, scalability, computational efficiency, etc.
%In view of the authors, emerging physics-informed techniques under certain Koopman-based model structures are promising to mitigate these challenges simultaneously. However, a few scientific issues remain to be solved.

\color{black}

\subsubsection{Linear Data-Driven Identification and Control}
Linear data-driven identification and control are still trending in certain power grid applications when the introduced model uncertainty is reasonably bounded (i.e., the small signal assumption roughly holds). A typical example is oscillation damping control as discussed in Section 5.1. %The reason is the inherent advantages of linear systems, which are hereby summarized:
%(i) their parameters correspond to physically interpretable parameters such as power flow Jacobians,
%\color{red} I think this statement is not true for most of linear methods except OU (yeah, so just deleted it) \color{black} 
%(i) their parameters can be identified optimally with the mature theory foundations of linear systems; (ii) it is easy to realize adaptiveness due to the parameter identification efficiency in linear models; (iii) linear models are capable enough and intuitive to model oscillatory modes in conventional power grids, and generally black-box linear models tend to have lower model variance than black-box nonlinear models; (iv) the linear model is intrinsically scalable due to the superposition and homogeneity properties.
%(v) adaptive linear control can be used to address nonlinear dynamics when the assumption holds that the dynamics evolving between adjacent time steps can be well approximated by a linearized model.
%\color{red} %check if it is repetitive (maybe OK based on current observation. The potential repetition is at the end of Section 3 whereas more comprehensive) 
%(i)(ii) were mentioned in Section 3. (iii) and (iv) were also mentioned somewhere else or intuitive. But more importantly, we want to talk about the development trend of linear data driven methods (how are they going to evolve? combining with online learning and robust control, it may address the challenges of uncertainty and nonlinearity? Or any other directions?). I think the discussion focus is not why linear data driven methods are trending. (modified)
%\color{black}
However, with the increasingly high penetration of volatile renewables and power electronics devices, the nonlinearity makes the effectiveness of linear models compromised. Nevertheless, adaptive linear control with online identification has the potential to address this issue to some extent by compensating for time-varying model uncertainty owing to the high learning efficiency and fewer data requirements of linear models, while needing further investigation to consolidate in practical applications.
%Even for grid applications where nonlinearity may become dominant (such as microgrid control and demand response discussed in Section 5.1), the linear model and identification can be seamlessly integrated with Koopman-based frameworks, and thus promising to address the nonlinearity in an adaptive Koopman-based fashion. \color{red} is the tone too high? Why do we focus on nonlinear methods then? \color{black}

Data-enabled predictive control (DeePC), a direct linear control method, is also emerging in control community and deserves attention for power grid applications. It is an efficient direct data-driven control method based on behavioral systems theory to learn a non-parametric system model that synthesizes the optimality of both identification and control\cite{Coulson2019}. The direct equivalence between DeePC and subspace predictive control has been demonstrated \cite{van2022data}. Adding quadratic regularization terms with DeePC leads to more robust identification against data noise \cite{huang2021quadratic}.  Although DeePC is still in the early stage, the linear nature is of potential to combine with online learning to adaptively address time-varying uncertainty and adopt Koopman operator-based structures to address nonlinearity. Also, direct optimal identification and control with a small piece of data subject to disturbances and noises may be further developed.

%\color{red} I think you may think about the trends of development of linear methods. For example, combining with online learning and robust control, it may address the challenges of uncertainty and nonlinearity? Also, direct linear identification and control is still at early stage, but may be further developed?....(done) 
\color{black}

%\color{black}
\subsubsection{Machine Learning-Based Methods}
%Power grids transmit and distribute electricity to vital end-users in societies; thus 
It is vital in real-world applications to ensure safe operation after applying any designed controllers. Generally speaking, most of the existing machine learning methods are based on universal approximators but the learning space for parametric training is too broad. Therefore, training of these universal learning machines is often mathematically (sub)optimal for training data while not being reasonable/generalizable for unforeseen cases in real-world applications. This tends to cause over-fitting and compromise the modeling reliability for safety-critical power grids.
\color{black} Among different machine learning-based methods discussed in Section 4, RL-based methods are the most popular nowadays and seem trending as they are where machine learning meets the feedback control. However, the time-consuming training (e.g., tens of thousands of iterations to converge \cite{Chen2022_RLreviewPS}) compromises the fast deployment and adaption to time-varying operating conditions and grid topologies. From the cyberinfrastructure perspective, their practical applications rely on data platforms with sufficient computing resources, data repositories, communication and appropriate technologies to improve training efficiency. From the power grid operation perspective, most of RLs are ``black-box" without enough physical interpretability and are not tractable to facilitate the incorporation of physics-inspired information. The safety and scalability of the power grid applications are problematic too. %In short, one may consider leveraging domain knowledge and exploiting application-specific structures to design tailored machine learning in the future as it may help achieve superior performance \cite{Chen2022_RLreviewPS}. 
%Koopman-based methods are more convenient to leverage them due to the two above-mentioned facts and thus by nature more interpretable and scalable.
\color{black}

To further enhance the RL reliability for increasingly complex power grids, safe exploration methods in RL can be considered to bound the parametric learning within a safety region \cite{garcia2015comprehensive}. Another possible way is to add safety constraints at the learned control policy in RL by either incorporating constraints \cite{li2019constrained} or adding a heuristic safety layer to adjust  the control actions \cite{Chen2022_RLreviewPS}. 
In addition to directly adding the safety constraints/layers in ``black-box" RL framework, the power grid modeling and control can be enhanced in a more interpretable framework by incorporating physics-inspired information in RL. For example, the Markov decision processes can be constrained with Bayesian to include physical priors through value or policy functions \cite{ghavamzadeh2015bayesian,brunke2022safe}. Additionally, model predictive control can be combined with RL, whereby the model information can be better incorporated into RL frameworks when applicable. For example, the authors in \cite{Li2022MPCandDP} leverage model predictive control (MPC) and an RL paradigm (dynamic programming), whereby more information can be processed effectively with enhanced reliability, and fast convergence in learning can be achieved.
In addition, the standard nonlinear model predictive control (NMPC) scheme can be used as a function approximator in RL, %whereby
through which the RL can benefit from the rich theory of NMPC with enhanced interpretability on closed-loop performance \cite{Gros2020}. Although these methods have yet to be fully investigated for power system applications, they are potential to data-driven model reliability and thus help augment grid operation safety. 
Some other technical trends in RL research community for more reliable learning also worth attention, including but not limited to the interpretable RL for enhanced consistency in safety-critical applications \cite{verma2018programmatically}, transfer RL\cite{parisotto2015actor} and meta RL for improving adaptiveness to new situations \cite{wang2016learning}, federated RL to preserve data-privacy \cite{qi2021federated}, Bayesian RL to incorporate uncertainty and prior knowledge in learning \cite{ghavamzadeh2015bayesian}, inverse RL to extract proper reward functions that the RL agents seek to maximize \cite{ng2000algorithms}, integral RL for continuous-time dynamics modeling \cite{abouheaf2019load}, etc.

%\color{red} I don't see why this paragraph is needed here. Is it asked by reviewers? Part of the contents (shortened one) may be added to the first paragraph of 5.4.2. (structure modified)
%\color{black} In summary, RL-based methods are popular nowadays as they are where machine learning meets the feedback control. From the general algorithm perspective, the learning process is nonlinear and requires large volumes of data to be shuffled to enhance the generalization capacity of learning. The time-consuming training of RL 
%(e.g., tens of thousands of iterations to converge \cite{Chen2022_RLreviewPS}) 
%due to large data volumes and high computational costs to solve nonlinear optimization also compromises the fast adaption to time-varying operating conditions and grid topologies. In other words, it requires sufficient computing resources and communication infrastructures, as well as technologies to improve training efficiency. Besides, most of RLs are ``black-box" without enough physical interpretability and are not tractable to facilitate the incorporation of physics-inspired information. The safety and scalability of the power grid applications are problematic. As suggested in \cite{Chen2022_RLreviewPS},  one may consider leveraging domain knowledge and exploiting application-specific structures to design tailored methods in the future as it can help achieve superior performance. 
%Koopman-based methods are more convenient to leverage them due to the two above-mentioned facts and thus by nature more interpretable and scalable.
\color{black}

Supervised learning-based surrogate models may also be designed for modern power grid control systems with physical information incorporated by unsupervised learning. The optimization objectives in such a learning paradigm are \cite{le2018supervised}\color{black}: (a) to minimize the training error of supervised learning, and meanwhile (b) to minimize a certain physics-informed error of unsupervised learning to regularize the supervised learning in (a). By incorporating regularization through unsupervised learning rather than directly adding regularization terms in objective functions, the generalization of learning machines can be improved due to the powerful unsupervised learning capacity and physics-informed shrink of parameter searching space. Similar ideas can be found in semi-supervised learning for power flow analysis in \cite{Hu2021}, where auto-encoders (i.e., a decoder + an encoder) are used to incorporate supervised and unsupervised learning. The encoder is where one can include physical knowledge of power grids such as the power mismatches \cite{baghaee2017three}, system topology information, etc \cite{Hu2021}. Besides, not only the constraints of systemwide power flow but also that of local DERs can be included during the model design and training phases such that more interpretable machine learning-based methods can be realized.

\subsubsection{Koopman-Based Methods and Others}
The above physics-inspired modeling techniques 
%in power grids reside in ``blackbox" machine learning. They 
are often trained with adequate offline datasets while rarely used for online modeling and real-time control. Thus, they may lack fast adaptiveness to time-varying operating conditions of modern power grids. In addition, the identification using the closed-loop control data is also challenging due to the time-varying uncertainty from control input channels.  %The incorporation of external control inputs is challenging too, because the ``black-box" modeling is not fully ``trustable" in nature especially for fast safety-critical scenarios such as primary and secondary large-disturbance stability and control in uncertain, nonlinear and low-inertia grid environments. 
In view of the authors, a dynamics-interpretable Koopman-inspired structure (e.g., Koopman operators and their variants such as EDMD and SINDy) with the help of emerging physics-informed techniques are promising for  online learning regarding adaptiveness and scalability. \color{black}Specifically, the Koopman observables $\Phi(\bm{x},\bm{u})$ in (\ref{eq:koc}) or the basis library $\bm\Theta$ in (\ref{eq:sindy}) define the Koopman-based model structure, which is the prerequisite for achieving certain physics interpretability on system dynamics and thus improving data-driven model efficacy and transparency for data-driven dynamical control. 

\color{black}Although promising, Koopman-based methods are still in the early stages of development. One of the main challenges limiting the applications of Koopman-based methods into power grids is the difficulty in determining the Koopman basis (i.e., Koopman observables or eigenfunctions) that formulates a proper Koopman state space structure. 
\color{black} By and large\color{black}, the Koopman basis is empirically determined based on power grid domain knowledge or learned directly from data with either universal training or self-dictation based on the dynamic relationship of the Koopman-based space and the original space \cite{lusch_deep_2018,Morton2019,Han2020,chen_variants_2012,Yeung2019}. \color{black}
For example, in high-inertia power grids dominated by multiple machines, the system dynamics are mainly driven by electromechanics, %and the interactions among different control loops and hierarchies are not significantly overlapping due to the high inertia of synchronous generators. In such cases, 
and the candidates of Koopman basis functions can be relatively straightforward (the combination of the trigonometric functions \cite{KORDA2018297,Netto2021}) by considering the differential equations of the electromechanical physical models and the network power flow. %The dominant Koopman observables then can be extracted from the basis candidates with data-driven identification. %The combination of trigonometric functions can capture the dominant dynamical characteristics, although it does not guarantee optimal design [11, 134]. %By observing the corresponding differential equations of the eletro-mechanic physical models with network power flow, the candidates of Koopman state space are straightforward (the combination of the trigonometric functions) \cite{KORDA2018297,Netto2021}. 
Although no guarantee for optimal design, the dominant dynamical characteristics can be well-captured. However, 
in future power grids with greater modernization, the determination of Koopman state space merely based on observation may not always work due to complex dynamics resulting from low inertia, a large number of diverse distributed energy resources, varying operating conditions as well as fast-evolving network topologies. The advancements of machine learning techniques now and future will offer tremendous opportunities to learn optimal Koopman embedding mapping. 
For example, a potential solution is to apply Koopman-based modeling in the RL frameworks, in which the Koopman-based methods can help RL enhance environment/value/policy function modeling and the RL framework in turn can help Koopman-based methods explore the problem inherent structure by pursuing Bellman optimality \cite{haykin_neural_2010}. Besides, 
physics-inspired domain information in power grids has yet to be incorporated in the deep learning-based Koopman basis discovery while deserving more attention. 
\color{black} Given appropriate Koopman structures, the development of Koopman-based online estimation and control are also trending because of their structural advantages inherited from linear systems.  
First, \color{black} as Koopman operator-based space is linear, mature linear identification techniques are applicable to enhance identification, which could be used for power grid applications in the future. For example, the linear matrix inequality (LMI) can be used to relax the minimization of the loss function with regularization terms (i.e., (\ref{eq:phyinformL}) or (\ref{eq:phyinform})) to convex problems \cite{Dahdah2021,James2021reg}, which can help realize modular, optimal and fast identification. 
\color{black}Besides, the mechanism with recursive measurement data may be designed to improve Koopman-based identification efficiency aiming at faster adaption to time-varying uncertainty and nonlinearity.
Second, \color{black} online ensemble learning to combine individual models is another way to further improve identification. For example, a ``soft switching" mechanism \cite{Zhang2010} can be employed to weigh linear and nonlinear models \cite{Ma2021}, and an online iterative ensemble structure\cite{Gong2022} can be used to combine small-signal physical models and large-signal models. Nonetheless, the stability and optimality of online ensemble control remain to be investigated. 
\color{black}Third\color{black}, the Koopman-enabled linear structure provides opportunities for the applications of distributed and cooperative MPC \cite{christofides2013distributed,negenborn2014distributed,stewart2010cooperative}, which are mature control techniques with well-characterized stability and robustness properties. This shed light on potential practical solutions to scalable integration of distributed energy resources into modern power grids.

\color{black}

%Data-enabled predictive control (DeePC) is another emerging data-driven method in control community and deserves attention for power grid applications. It is an efficient direct data-driven control method based on behavioural systems theory to learn a non-parametric system model with online adaptiveness \cite{Coulson2019}. The direct equivalence between DeePC and subspace predictive control has been demonstrated \cite{van2022data}. Adding quadratic regularization term with DeePC leads to more robust identification against data noise \cite{huang2021quadratic}.  Although DeePC is linear, it is potential to address nonlinearity due to the online adaptiveness and robustness. It is also potential to combine DeePC with Koopman operators to address the nonlinearity in power grids. 
%\color{red} you may consider moving part of this paragraph to 5.3.1 (DeePC part has been moved to the end of Section 5.3.1)\color{black}

\subsubsection{Data Sources}
%\color{red} My understanding is OPAL-RT or hard-ware-in-the-loop simulation is digital twin for power systems. Are you sure there are no existing machine-learning based works using real-time simulations for validation? (I think DT is a self-adaptive virtual model which can be used to fully observe and predict the real system and could be further used for control in real-time. So it is more than a simulator. While HIL or real-time simulation can facilitate the realization of a DT)\color{black}

To the best of our knowledge, almost all the existing RL-based control are based on power system simulators. The supervised learning-based modeling and Koopman-based modeling for control are based on simulation data, historical experimental offline data or online data. 
%\color{black} From the general algorithm perspective, the learning process is nonlinear and requires a large number of simulations, with data being shuffled to enhance the generalization capacity of learning. The training of RL is usually time-consuming (e.g., tens of thousands of iterations to converge \cite{Chen2022_RLreviewPS}), also posing challenges to online adaptiveness.
%\color{red}the previous sentences highlighted in blue don't closely related to what is discussed next and may be deleted? (commented out)
\color{black}
Real-time simulation is an emerging trend that enables the development of high-fidelity simulators for complex modern power systems, offering the potential for future implementation of ``digital twins" for online monitoring and control. %to help realize complex modern power system simulation models of high fidelity, and paves the road to realizing ``digital twin" in the future for online monitoring and control. 
The digital twin is an up-to-date representation of an actual current asset in operation that includes the asset’s condition and relevant historical data \cite{he2019preliminary}. The real-time simulator and ultimately the digital twin can be used as the simulator for RL and the full-state data generator for supervised learning. \color{black}
However, it's important to note that even with high-fidelity simulators or experimental data, there may exist a small gap between these representations and the  real-world systems, leading to uncertainty propagation and thus compromising control performance. %\color{black} Even though, a tiny gap between simulators/experimental data and real-world underlying systems may propagate uncertainty and thus compromise control performance. 
To mitigate this, some countermeasures may be taken such as robust (adversarial) RL methods in simulator-based policy training \cite{Chen2022_RLreviewPS,pinto2017robust,morimoto2005robust}, and data augmentation techniques, such as adversarial NNs \cite{antoniou2017data}, can generate sufficient training data that includes necessary information for robust supervised learning-based modeling for control. %and data augmentation techniques such as adversarial NNs \cite{antoniou2017data} can be used to generate adequate training data that includes sufficient information for supervised learning-based modeling for control. 
\color{black}

\color{black}

\section{Conclusion}
This paper provides a comprehensive review on emerging data-driven control for the applications in modern power grids. The data-driven control mainly consists of two ingredients --- identification and control, which should be conducted jointly in a either sequential or simultaneous manner. To realize real-world applications for safety-critical power grids of high complexity, physics-informed methodology is necessary while it remains to be studied to realize a practical solution that is theoretically sound with performance guarantee. In general, one needs to consider the physical constraints and find elegant ways to incorporate them in identification and control. To this end, 
%In general, one needs to consider the physical constraints and find elegant ways to incorporate them in identification and control. 
%
combining learning-based methods, physical domain knowledge, and the Koopman-based model structure seems promising as mature linear system control methods with well-characterized stability and robustness can be readily applied to the linear Koopman-based model. 
%properties can be applied to the linear Koopman-based model. Besides, enabled by the Koopman operator theory may . 
Besides, various machine learning methods including supervised, unsupervised learning and RL
may be used jointly % Combining different learning-based methods under the Koopman-based model structure is promising. Although pure machine learning methods including supervised, unsupervised learning and RL are ``black-box" in nature, they are powerful approximators that can be used jointly 
in exploring and exploiting model structure and parameters with proper physics-informed learning objectives and constraints defined by users.

%\color{black}

%\appendices
%\section{} 

%\label{Aderivation}

\bibliographystyle{elsarticle-num}
\bibliography{mybib.bib}

%\bibitem{Bidram:2013}
%A. Bidram, A. Davoudi, F. L. Lewis, and J. M. Guerrero, {\em Distributed cooperative secondary control of microgrids using feedback linearization control}.
%\newblock IEEE Trans. Power Syst., vol. 28, no. 3, pp. 3462–3470, 2013.

\end{document}